\documentclass[fleqn,usenatbib]{mnras}

\usepackage{newtxtext,newtxmath}
\usepackage[T1]{fontenc}
\usepackage{ae,aecompl}

\usepackage{amsmath,amsfonts,bm}   
\usepackage{mathrsfs}    
\usepackage{graphicx}
\usepackage[utf8]{inputenc}
\usepackage{color}
\usepackage{adjustbox}   
\usepackage{hyperref}    

\newcommand{\tild}{~}
\newcommand{\si}{$\sigma_e$\tild}
\newcommand{\sinosp}{$\sigma_e$}
\newcommand{\simath}{\sigma_e}

\newcommand{\mbh}{$M_{BH}$\tild}
\newcommand{\mbhnosp}{$M_{BH}$}
\newcommand{\mbhmath}{M_{BH}}

\newcommand{\lk}{$L_K$\tild}

\newcommand{\mbul}{$M_{bul}$\tild}

\newcommand{\mbulmath}{M_{bul}}

\newcommand{\re}{$R_e$\tild}
\newcommand{\renosp}{$R_e$}
\newcommand{\remath}{R_e}
\newcommand{\Hop}{$M_{Hop}$\tild}
\newcommand{\mvir}{$M_{vir}$\tild}
\newcommand{\mgrav}{$U_{grav}$\tild}

\newcommand{\Spitzer}{\textit{Spitzer}\tild}
\newcommand{\htw}{\hline \hline}
\newcommand{\hdr}{\htw \hline}

\title[]{The Fundamental Relation between Supermassive Black Holes and Their Host Galaxies}

\author[S. de Nicola et al.]{
Stefano de Nicola,$^{1,2}$\thanks{E-mail: st.denicola2@gmail.com}
Alessandro Marconi,$^{3,4}$
and Giuseppe Longo$^{1,5,6}$
\\
$^1$ Università degli Studi di Napoli ``Federico II'', Department of Physics ``E. Pancini'', Via Cinthia 6, 80126 Napoli, Italy\\
$^2$ Max-Planck-Institut f{\"u}r extraterrestrische Physik, Giessenbachstrasse 1, D-85748 Garching, Germany\\
$^3$ Università degli Studi di Firenze, Department of Physics and Astronomy, Via G. Sansone 1, 50019 Sesto Fiorentino, Firenze, Italy\\
$^4$ INAF–Osservatorio Astrofisico di Arcetri, Largo Enrico Fermi 5, I-50125, Firenze, Italy\\
$^5$ INFN, Napoli Unit, via Cinthia 6, 80126, Napoli, Italy\\
$^6$ Merle Kingsley Distinguished Visitor, CALTECH, Pasadena 90125-CA, USA}

\date{Accepted XXX. Received YYY; in original form ZZZ}

\pubyear{2019}

\begin{document}
\label{firstpage}
\pagerange{\pageref{firstpage}--\pageref{lastpage}}
\maketitle

\begin{abstract}
We study the correlations between Supermassive Black Holes (BH) and their host galaxies, using a sample of 83 BH masses collected from the most recent and reliable spatially resolved estimates available from the literature. We confirm the mono- and bivariate correlations between SMBHs and the bulges of their host galaxies, confirming that the correlation with the effective velocity dispersion is not significantly improved by higher dimensionality. Instead, pseudobulges do not seem to correlate with their SMBHs, probably because their secular evolution is often unable to trigger accretion onto the central BH. 
We then present a novel approach aimed at finding the fundamental relation between SMBHs and their host galaxies. For the first time, we analytically combine BH masses with the Fundamental Plane (FP), showing that \mbhnosp-\si appears to be the fundamental relation rather than a putative ``BH Fundamental Plane'' of the kind \mbhnosp-\sinosp-\renosp. These results can be explained by a picture which sees the \mbhnosp-\si relation as a natural outcome of the change in AGN feedback from momentum- to energy-driven. The other scaling relations are then established through the FP.

\end{abstract}




\section{Introduction}

The studies conducted in the last 25 years have shown that Supermassive Black Holes (hereafter SMBHs) play a crucial role in the formation and the evolution of their host spheroids (see \citealt{K13} and \citealt{Graham16} for reviews). The most significant pieces of evidence are given by the correlations between the BH mass \mbh and the effective velocity dispersion \si \citep{Ferrarese00, Gebhardt00, Tremaine02, Gultekin09b} and that between \mbh and the bulge mass \mbul \citep{Magorrian98, Marconi03, Haering04} where, in the case of elliptical galaxies, the bulge corresponds to the whole spheroid. Other monovariate correlations which have been investigated are those with the bulge kinetic energy $\mbulmath \simath^2$ \citep{Mancini12,Feoli09}, the Dark Matter (DM) halo (\citealt{Ferrarese02}, but see \citealt{Kormendy11b}), the Sérsic index \citep{Savorgnan13}, the pitch angle \citep{Davis17} or that with the core radius \citep{S16}.\\
\indent Recently, there have been studies investigating whether higher dimensionality, i.e. a relation combining \mbh with multiple galaxy parameters, can yield better \mbh predictions (e.g. \citealt{Hopkins07a, Hopkins07b, Sani11, Beifiori12, S16}). Finding the fundamental relation between SMBHs and their hosts is of great importance since it would shed light on the physical mechanism behind these correlations and would provide us with the parameter(s) which yield(s) the most accurate \mbh predictions. Works focusing on BH-galaxy scaling relations have shown that the \mbhnosp-\si relation has the lowest intrinsic scatter (e.g. \citealt{Gultekin09b,S16,VDB16}) and is just marginally (if at all) improved by higher dimensionality \citep{Beifiori12, S16}. In a recent review, \citet{King15} show how this relation can be explained by a change in AGN feedback from momentum-driven to energy-driven and how it could generate the canonical $\mbhmath\propto\mbulmath^1$. Instead, in \citet{Hopkins07b} it is found that the fundamental relation should be a plane of the kind $\mbhmath\propto\simath^\alpha\remath^\beta$ at 3$\sigma$ significance level. A total of 5 BH-galaxy bivariate correlations are detected in the very exhaustive study of \citet{S16}, although \citet{Shankar16, Shankar17, Shankar19} have raised the issue of a presence of a bias in favour of more massive BHs in their sample.\\
\indent The most general description of a bulge is given by the Fundamental Plane (hereafter FP, \citealt{Djorgovski87}). This plane is given by the combination of the virial theorem and a tilt mostly given by the weak dependence of the mass-to-light ratio $M/L$ on $L$ itself \citep{Cappellari06}\footnote{The SMBH itself could contribute to this tilt, as we show in Sec.\tild\ref{modellino}.}. Thus, in order to have the most general picture of this BH-bulge interaction, the whole FP should be combined with BH masses, also because the BH itself is part of the system probed by the FP. So far, there has been only one study \citep{VDB16} focusing on unifying the FP with BH masses. In that work, the author shows that a relation of the form $M_{BH} \propto \left(\frac{L_K}{R_e}\right)^{3.8}$ should be used to measure BH masses when \si measurements are not available\footnote{Note that in this work the author focuses on \textit{whole} galaxies rather than on bulges solely.}.\\
\indent Besides classical bulges, we also encounter the so-called \textit{pseudobulges} \citep{Kormendy04, Fisher08, Fisher13}. These systems actively form stars and are rotationally supported, thus resembling disks more than classical bulges. Our own Galaxy is the closest example of pseudobulge \citep{K13}. Such systems do not seem to follow the BH-hosts scaling relations \citep{Graham13, S16} and may not even lie on the FP (\citealt{Kormendy04}, but see Tab. 10 of \citealt{S16}), although these are quite difficult to identify \citep{Graham14}.\\
\indent In this work, we analyze these existing correlations and propose a novel multivariate analytic approach aimed at combining BH masses with the FP also taking into account covariances and correlations between observables. This enables us to verify, among all scaling relations and regardless of the dimensionality, which one yields the best predictions of the others. The paper is structured as follows. In Section 2, we present the dataset. In Section 3, we fit linear regressions to our data. In Section 4, we present our approach aimed at unifying the FP and BH masses. In Section 5, we briefly examine the causality of our results and draw our conclusions. Notes on galaxies omitted from the analysis can be found in App.\tild\ref{Appendice}. Unless differently specified, we will always provide $1\sigma$ uncertainties on all variables.

\begin{table*}  
	\begin{tabular}{lcccccccc}
		\hdr 
		Galaxy & Morphology & A & Distance & \mbh & \si & \lk & \re & B \\ 
        & & & (Mpc) & ($\log\,M_\odot$) & ($\log\,km/s$) & ($\log\,L_\odot$) & ($\log\,kpc$) & \\
		\hline
		Circinus & SABb: &   3 & 2.82 $\pm$ 0.47 & 6.23 $\pm$ 0.10 & 1.90 $\pm$ 0.02 & 10 $\pm$ 0.12 & -0.91 $\pm$ 0.07 &   1 \\ 
		A1836 & BCGE &   0 & 152.4 $\pm$ 8.4 & 9.57 $\pm$ 0.06 & 2.46 $\pm$ 0.02 & 11.75 $\pm$ 0.06 & 0.89 $\pm$ 0.02 &   0 \\ 
		IC1459 & E4 &   0 & 28.9 $\pm$ 3.7 & 9.39 $\pm$ 0.08 & 2.52 $\pm$ 0.01 & 11.70 $\pm$ 0.06 & 0.90 $\pm$ 0.06 &   1 \\ 
		NGC524 & S0 &   2 & 24.2 $\pm$ 2.2 & 8.94 $\pm$ 0.05 & 2.39 $\pm$ 0.02 & 10.52 $\pm$ 0.08 & 0.17 $\pm$ 0.07 &   1 \\ 
		NGC821 & S0 &   1 & 23.4 $\pm$ 1.8 & 8.22 $\pm$ 0.21 & 2.32 $\pm$ 0.02 & 10.84 $\pm$ 0.31 & 0.33 $\pm$ 0.03 &   1 \\ 
		NGC1023 & SB0 &   2 & 10.81 $\pm$ 0.80 & 7.62 $\pm$ 0.06 & 2.31 $\pm$ 0.02 & 10.45 $\pm$ 0.07 & -0.41 $\pm$ 0.03 &   1 \\ 
		NGC1399 & E1 &   0 & 20.85 $\pm$ 0.67 & 8.95 $\pm$ 0.31 & 2.498 $\pm$ 0.004 & 11.81 $\pm$ 0.06 & 1.53 $\pm$ 0.01 &   1 \\ 
		NGC2273 & SBa &   3 & 29.5 $\pm$ 1.9 & 6.93 $\pm$ 0.04 & 2.10 $\pm$ 0.03 & 10.43 $\pm$ 0.40 & -0.57 $\pm$ 0.03 &   1 \\ 
		NGC2549 & S0/ &   2 & 12.7 $\pm$ 1.6 & 7.16 $\pm$ 0.37 & 2.16 $\pm$ 0.02 & 9.73 $\pm$ 0.06 & -0.72 $\pm$ 0.06 &   1 \\ 
		NGC3115 & S0/ &   2 & 9.54 $\pm$ 0.40 & 8.95 $\pm$ 0.10 & 2.36 $\pm$ 0.02 & 10.93 $\pm$ 0.06 & 0.20 $\pm$ 0.06 &   1 \\ 
		NGC3227 & SBa &   3 & 23.7 $\pm$ 2.6 & 7.32 $\pm$ 0.23 & 2.12 $\pm$ 0.04 & 9.93 $\pm$ 0.25 & -0.28 $\pm$ 0.05 &   1 \\ 
		NGC3245 & S0 &   2 & 21.38 $\pm$ 1.97 & 8.38 $\pm$ 0.11 & 2.31 $\pm$ 0.02 & 10.20 $\pm$ 0.06 & -0.60 $\pm$ 0.04 &   1 \\ 
		NGC3377 & E5 &   1 & 10.99 $\pm$ 0.46 & 8.25 $\pm$ 0.25 & 2.16 $\pm$ 0.02 & 10.64 $\pm$ 0.25 & 0.52 $\pm$ 0.02 &   1 \\ 
		NGC3384 & SB0 &   3 & 11.49 $\pm$ 0.74 & 7.03 $\pm$ 0.21 & 2.16 $\pm$ 0.02 & 10.20 $\pm$ 0.06 & -0.51 $\pm$ 0.03 &   1 \\ 
		NGC3393 & SABa &   3 & 49.2 $\pm$ 8.2 & 7.20 $\pm$ 0.33 & 2.17 $\pm$ 0.03 & 10.62 $\pm$ 0.25 & -0.48 $\pm$ 0.07 &   1 \\ 
		NGC3585 & S0 &   2 & 20.5 $\pm$ 1.7 & 8.52 $\pm$ 0.13 & 2.33 $\pm$ 0.02 & 11.45 $\pm$ 0.25 & 0.93 $\pm$ 0.07 &   1 \\ 
		NGC3608 & E1 &   0 & 22.8 $\pm$ 1.5 & 8.67 $\pm$ 0.10 & 2.26 $\pm$ 0.02 & 11.04 $\pm$ 0.25 & 0.68 $\pm$ 0.03 &   1 \\ 
		NGC3842 & E1 &   0 & 92 $\pm$ 11 & 9.96 $\pm$ 0.14 & 2.43 $\pm$ 0.04 & 12.04 $\pm$ 0.06 & 1.52 $\pm$ 0.05 &   1 \\ 
		NGC3998 & S0 &   2 & 14.3 $\pm$ 1.2 & 8.93 $\pm$ 0.05 & 2.44 $\pm$ 0.01 & 10.15 $\pm$ 0.31 & -0.48 $\pm$ 0.04 &   1 \\ 
		NGC4026 & S0 &   2 & 13.3 $\pm$ 1.7 & 8.26 $\pm$ 0.12 & 2.25 $\pm$ 0.02 & 9.86 $\pm$ 0.31 & -0.39 $\pm$ 0.06 &   1 \\ 
		NGC4258 & SABbc &   2 & 7.27 $\pm$ 0.50 & 7.58 $\pm$ 0.03 & 2.06 $\pm$ 0.04 & 10.03 $\pm$ 0.03 & -0.33 $\pm$ 0.03 &   1 \\ 
		NGC4261 & E2 &   0 & 32.4 $\pm$ 2.8 & 8.72 $\pm$ 0.10 & 2.5 $\pm$ 0.02 & 11.53 $\pm$ 0.25 & 0.87 $\pm$ 0.04 &   1 \\ 
		NGC4291 & E2 &   0 & 26.6 $\pm$ 3.9 & 8.99 $\pm$ 0.16 & 2.38 $\pm$ 0.02 & 10.86 $\pm$ 0.25 & 0.30 $\pm$ 0.06 &   1 \\ 
		NGC4459 & E2 &   1 & 16.01 $\pm$ 0.52 & 7.84 $\pm$ 0.09 & 2.22 $\pm$ 0.02 & 10.64 $\pm$ 0.25 & 0.00 $\pm$ 0.01 &   1 \\ 
		NGC4473 & E5 &   1 & 15.25 $\pm$ 0.49 & 7.95 $\pm$ 0.24 & 2.28 $\pm$ 0.02 & 10.80 $\pm$ 0.25 & 0.44 $\pm$ 0.01 &   1 \\ 
		NGC4564 & S0 &   2 & 15.94 $\pm$ 0.51 & 7.95 $\pm$ 0.12 & 2.21 $\pm$ 0.02 & 10.15 $\pm$ 0.06 & -0.41 $\pm$ 0.01 &   1 \\ 
		NGC4596 & SB0 &   2 & 16.5 $\pm$ 6.2 & 7.88 $\pm$ 0.26 & 2.13 $\pm$ 0.02 & 10.34 $\pm$ 0.06 & -0.14 $\pm$ 0.16 &   1 \\ 
		NGC4649 & E2 &   0 & 16.46 $\pm$ 0.61 & 9.67 $\pm$ 0.10 & 2.58 $\pm$ 0.02 & 11.66 $\pm$ 0.06 & 0.90 $\pm$ 0.02 &   0 \\ 
		NGC4697 & E5 &   1 & 12.54 $\pm$ 0.40 & 8.13 $\pm$ 0.01 & 2.25 $\pm$ 0.02 & 11.17 $\pm$ 0.31 & 0.64 $\pm$ 0.01 &   1 \\ 
		NGC4889 & E4 &   0 & 102.0 $\pm$ 5.1 & 10.32 $\pm$ 0.44 & 2.54 $\pm$ 0.01 & 12.25 $\pm$ 0.06 & 1.47 $\pm$ 0.02 &   1 \\ 
		NGC5077 & E3 &   0 & 38.7 $\pm$ 8.4 & 8.93 $\pm$ 0.27 & 2.35 $\pm$ 0.02 & 11.42 $\pm$ 0.06 & 0.64 $\pm$ 0.09 &   1 \\ 
		NGC5128 & E &   0 & 3.62 $\pm$ 0.20 & 7.75 $\pm$ 0.08 & 2.18 $\pm$ 0.02 & 10.80 $\pm$ 0.31 & 0.03 $\pm$ 0.02 &   1 \\ 
		NGC5576 & E3 &   1 & 25.7 $\pm$ 1.7 & 8.44 $\pm$ 0.13 & 2.26 $\pm$ 0.02 & 11.02 $\pm$ 0.06 & 0.79 $\pm$ 0.03 &   1 \\ 
		NGC5845 & E3 &   1 & 25.9 $\pm$ 4.1 & 8.69 $\pm$ 0.16 & 2.38 $\pm$ 0.02 & 10.43 $\pm$ 0.31 & -0.41 $\pm$ 0.07 &   1 \\ 
		NGC6086 & E &   0 & 138 $\pm$ 11 & 9.57 $\pm$ 0.17 & 2.500 $\pm$ 0.002 & 11.87 $\pm$ 0.08 & 1.20 $\pm$ 0.04 &   0 \\ 
		NGC6251 & E1 &   0 & 108.4 $\pm$ 9.0 & 8.79 $\pm$ 0.16 & 2.46 $\pm$ 0.02 & 11.94 $\pm$ 0.06 & 1.20 $\pm$ 0.04 &   1 \\ 
		NGC7052 & E3 &   0 & 70.4 $\pm$ 8.4 & 8.60 $\pm$ 0.23 & 2.42 $\pm$ 0.02 & 11.77 $\pm$ 0.06 & 1.10 $\pm$ 0.05 &   1 \\ 
		NGC7582 & SBab &   3 & 22.3 $\pm$ 9.8 & 7.74 $\pm$ 0.20 & 2.19 $\pm$ 0.05 & 10.61 $\pm$ 0.32 & -0.62 $\pm$ 0.19 &   0 \\ 
		NGC7768 & E4 &   0 & 116 $\pm$ 27 & 9.13 $\pm$ 0.18 & 2.41 $\pm$ 0.04 & 12.00 $\pm$ 0.25 & 1.37 $\pm$ 0.10 &   1 \\ 
		UGC3789 & SABab &   3 & 49.9 $\pm$ 5.4 & 6.99 $\pm$ 0.08 & 2.03 $\pm$ 0.05 & 10.33 $\pm$ 0.31 & -0.24 $\pm$ 0.05 &   1 \\ 
		NGC1332 & S0 &   2 & 22.3 $\pm$ 1.8 & 8.82 $\pm$ 0.10 & 2.47 $\pm$ 0.01 & 11.20 $\pm$ 0.31 & 0.29 $\pm$ 0.06 &   1 \\ 
		NGC1374 & E3 &   1 & 19.23 $\pm$ 0.66 & 8.76 $\pm$ 0.06 & 2.23 $\pm$ 0.01 & 10.72 $\pm$ 0.06 & 0.36 $\pm$ 0.01 &   1 \\ 
		NGC1407 & E0 &   0 & 28.0 $\pm$ 3.4 & 9.65 $\pm$ 0.08 & 2.442 $\pm$ 0.003 & 11.72 $\pm$ 0.12 & 0.97 $\pm$ 0.05 &   0 \\ 
		NGC1550 & SA0 &   0 & 51.6 $\pm$ 5.6 & 9.57 $\pm$ 0.07 & 2.44 $\pm$ 0.02 & 11.32 $\pm$ 0.10 & 0.66 $\pm$ 0.05 &   0 \\ 
		NGC3091 & E3 &   0 & 51.2 $\pm$ 8.3 & 9.56 $\pm$ 0.07 & 2.48 $\pm$ 0.02 & 11.75 $\pm$ 0.06 & 1.10 $\pm$ 0.07 &   1 \\ 
		NGC3368 & SABab &   3 & 10.40 $\pm$ 0.96 & 6.88 $\pm$ 0.08 & 2.122 $\pm$ 0.003 & 10.09 $\pm$ 0.06 & -0.57 $\pm$ 0.04 &   1 \\ 
		NGC3489 & SAB0 &   3 & 12.10 $\pm$ 0.84 & 6.78 $\pm$ 0.05 & 1.949 $\pm$ 0.002 & 9.68 $\pm$ 0.25 & -1.00 $\pm$ 0.03 &   1 \\ 
		NGC4751 & E &   1 & 26.9 $\pm$ 2.9 & 9.15 $\pm$ 0.06 & 2.56 $\pm$ 0.02 & 10.95 $\pm$ 0.09 & 0.52 $\pm$ 0.05 &   0 \\ 
		NGC5328 & E &   0 & 64.1 $\pm$ 7.0 & 9.67 $\pm$ 0.16 & 2.523 $\pm$ 0.002 & 11.71 $\pm$ 0.09 & 0.94 $\pm$ 0.05 &   0 \\ 
		NGC5516 & E &   0 & 58.4 $\pm$ 6.4 & 9.52 $\pm$ 0.06 & 2.52 $\pm$ 0.02 & 11.83 $\pm$ 0.09 & 1.30 $\pm$ 0.05 &   0 \\ 
		NGC6861 & E &   1 & 27.3 $\pm$ 4.6 & 9.30 $\pm$ 0.08 & 2.590 $\pm$ 0.003 & 11.14 $\pm$ 0.13 & 0.32 $\pm$ 0.07 &   0 \\ 
		NGC7619 & E &   0 & 51.5 $\pm$ 7.4 & 9.40 $\pm$ 0.11 & 2.47 $\pm$ 0.01 & 11.78 $\pm$ 0.25 & 1.16 $\pm$ 0.06 &   1 \\ 
		NGC2748 & Sc &   3 & 23.4 $\pm$ 8.2 & 7.65 $\pm$ 0.24 & 2.06 $\pm$ 0.02 & 9.84 $\pm$ 0.25 & -0.39 $\pm$ 0.15 &   1 \\ 
		NGC4151 & Sa &   2 & 20.0 $\pm$ 2.8 & 7.81 $\pm$ 0.08 & 2.19 $\pm$ 0.02 & 10.61 $\pm$ 0.25 & -0.18 $\pm$ 0.06 &   1 \\ 
		NGC7457 & S0 &   2 & 12.5 $\pm$ 1.2 & 6.95 $\pm$ 0.30 & 1.83 $\pm$ 0.02 & 9.69 $\pm$ 0.08 & -0.28 $\pm$ 0.04 &   1 \\ 
		NGC307 & S0 &   2 & 52.8 $\pm$ 5.7 & 8.60 $\pm$ 0.06 & 2.31 $\pm$ 0.01 & 10.50 $\pm$ 0.05 & -0.31 $\pm$ 0.05 &   0 \\ 
		NGC3627 & SAB(s)b &   3 & 10.0 $\pm$ 1.1 & 6.93 $\pm$ 0.05 & 2.088 $\pm$ 0.002 & 9.45 $\pm$ 0.09 & -1.08 $\pm$ 0.05 &   0 \\ 
		NGC3923 & E4 &   1 & 20.9 $\pm$ 2.7 & 9.45 $\pm$ 0.12 & 2.35 $\pm$ 0.02 & 11.50 $\pm$ 0.11 & 0.89 $\pm$ 0.06 &   0 \\ 
		NGC4486A & E2 &   1 & 16.00 $\pm$ 0.52 & 7.10 $\pm$ 0.15 & 2.16 $\pm$ 0.01 & 10.08 $\pm$ 0.05 & -0.19 $\pm$ 0.01 &   0 \\ 
		NGC4501 & SA(rs)b &   3 & 16.5 $\pm$ 1.1 & 7.30 $\pm$ 0.08 & 2.20 $\pm$ 0.01 & 10.16 $\pm$ 0.07 & -0.40 $\pm$ 0.03 &   0 \\ 
		NGC5018 & E3 &   1 & 40.6 $\pm$ 4.9 & 8.02 $\pm$ 0.08 & 2.32 $\pm$ 0.01 & 11.54 $\pm$ 0.09 & 0.62 $\pm$ 0.05 &   0 \\ 
		NGC5419 & E &   0 & 56.2 $\pm$ 6.1 & 9.86 $\pm$ 0.14 & 2.56 $\pm$ 0.01 & 12.00 $\pm$ 0.09 & 1.26 $\pm$ 0.05 &   0 \\ 
		IC4296 & BCGE &   0 & 49.2 $\pm$ 3.6 & 9.11 $\pm$ 0.07 & 2.51 $\pm$ 0.02 & 11.78 $\pm$ 0.25 & 1.21 $\pm$ 0.03 &   1 \\ 
		NGC1277 & S0/ &   2 & 73.0 $\pm$ 7.3 & 9.70 $\pm$ 0.05 & 2.52 $\pm$ 0.07 & 10.83 $\pm$ 0.08 & 0.09 $\pm$ 0.04 &   0 \\ 
		IC2560 & SBbc &   3 & 33.2 $\pm$ 3.3 & 6.59 $\pm$ 0.16 & 2.15 $\pm$ 0.02 & 10.13 $\pm$ 0.25 & -0.14 $\pm$ 0.04 &   1 \\ 
		\htw 
	\end{tabular}
    \label{par_table}
    \end{table*}

 \begin{table*}
 \begin{tabular}{lcccccccc}
		\hdr 
		Galaxy & Morphology & A & Distance & \mbh & \si & \lk & \re & B \\ 
        & & & (Mpc) & ($\log\,M_\odot$) & ($\log\,km/s$) & ($\log\,L_\odot$) & ($\log\,kpc$) & \\
		\hline
    NGC224 & Sb &   2 & 0.77 $\pm$ 0.03 & 8.15 $\pm$ 0.16 & 2.23 $\pm$ 0.02 & 10.34 $\pm$ 0.10 & -0.19 $\pm$ 0.02 &   1 \\ 
		NGC4472 & E2 &   0 & 17.14 $\pm$ 0.59 & 9.40 $\pm$ 0.04 & 2.48 $\pm$ 0.01 & 11.86 $\pm$ 0.06 & 1.05 $\pm$ 0.01 &   1 \\ 
		NGC3031 & Sb &   2 & 3.60 $\pm$ 0.13 & 7.81 $\pm$ 0.13 & 2.15 $\pm$ 0.02 & 10.43 $\pm$ 0.31 & -0.24 $\pm$ 0.02 &   1 \\ 
		NGC4374 & E1 &   0 & 18.51 $\pm$ 0.60 & 8.97 $\pm$ 0.05 & 2.47 $\pm$ 0.02 & 11.64 $\pm$ 0.25 & 1.07 $\pm$ 0.01 &   1 \\ 
		NGC4486 & E1 &   0 & 16.68 $\pm$ 0.62 & 9.68 $\pm$ 0.04 & 2.51 $\pm$ 0.03 & 11.64 $\pm$ 0.25 & 0.85 $\pm$ 0.02 &   1 \\ 
		NGC4594 & Sa &   2 & 9.87 $\pm$ 0.82 & 8.82 $\pm$ 0.04 & 2.38 $\pm$ 0.02 & 10.79 $\pm$ 0.25 & -0.03 $\pm$ 0.08 &   1 \\ 
		NGC3379 & E1 &   0 & 10.70 $\pm$ 0.54 & 8.62 $\pm$ 0.11 & 2.31 $\pm$ 0.02 & 10.96 $\pm$ 0.25 & 0.42 $\pm$ 0.02 &   1 \\ 
		NGC221 & E2 &   1 & 0.80 $\pm$ 0.03 & 6.39 $\pm$ 0.19 & 1.89 $\pm$ 0.02 & 9.12 $\pm$ 0.04 & -0.90 $\pm$ 0.02 &   0 \\ 
		CygnusA & E &   0 & 242 $\pm$ 24 & 9.42 $\pm$ 0.12 & 2.43 $\pm$ 0.05 & 12.19 $\pm$ 0.10 & 1.46 $\pm$ 0.04 &   0 \\
		
		NGC1271 & SB0 &   2 & 80.0 $\pm$ 8.0 & 9.48 $\pm$ 0.15 & 2.45 $\pm$ 0.01 & 11.07 $\pm$ 0.08 & 0.34 $\pm$ 0.07 &   0 \\ 
		NGC1275 & E &   1 & 73.8 $\pm$ 7.4 & 8.90 $\pm$ 0.24 & 2.39 $\pm$ 0.08 & 11.84 $\pm$ 0.08 & 1.15 $\pm$ 0.04 &   0 \\ 
		NGC1600 & E &   0 & 64.0 $\pm$ 6.4 & 10.23 $\pm$ 0.04 & 2.47 $\pm$ 0.02 & 11.86 $\pm$ 0.08 & 1.08 $\pm$ 0.04 &   0 \\ 
		NGC3706 & E &   0 & 46.0 $\pm$ 4.6 & 8.78 $\pm$ 0.06 & 2.51 $\pm$ 0.01 & 11.58 $\pm$ 0.08 & 0.80 $\pm$ 0.04 &   0 \\ 
		NGC5252 & S0 &   2 & 92.0 $\pm$ 9.2 & 8.98 $\pm$ 0.23 & 2.28 $\pm$ 0.02 & 11.49 $\pm$ 0.09 & 0.88 $\pm$ 0.06 &   0 \\ 
        NGC4339 & E &  1 & 16.0 $\pm$ 1.6 & 7.63 $\pm$ 0.36 & 1.98 $\pm$ 0.02 & 10.26 $\pm$ 0.25 & 0.37 $\pm$ 0.04 &   0 \\
        NGC4434 & E &  1 & 22.4 $\pm$ 2.2 & 7.85 $\pm$ 0.15 & 1.99 $\pm$ 0.02 & 10.28 $\pm$ 0.25 & 0.20 $\pm$ 0.04 &   0 \\
        NGC4578 & E &  1 & 16.3 $\pm$ 1.6 & 7.28 $\pm$ 0.22 & 2.03 $\pm$ 0.02 & 10.33 $\pm$ 0.25 & 0.49 $\pm$ 0.04 &   0 \\
        NGC4762 & E &  1 & 22.6 $\pm$ 2.3 & 7.36 $\pm$ 0.14 & 2.13 $\pm$ 0.02 & 11.05 $\pm$ 0.25 & 1.06 $\pm$ 0.04 &   0 \\
        \htw 
 \end{tabular}
\caption{\textit{Col.1}: Galaxy name. \textit{Col.2}: Morphology. \textit{Col.3}: Flag $A$, $A$=0 indicates core ellipticals, $A$=1 indicates power-law ellipticals, $A$=2 indicates bulges, $A$=3 indicates pseudobulges \citep{S16}. \textit{Col.4}: Distance. \textit{Col.5}: BH mass. \textit{Col.6}: Effective velocity dispersion. \textit{Col.7}: Luminosity, measured either at 3.6$\mu$m \citep{Savorgnan16b} or from 2MASS data (\citealt{VDB16}, §\tild\ref{gal_par}. \textit{Col.8}: Effective radius, coming from the same analysis as $L$. \textit{Col.9}: Flag $B$, $B$=0 indicates $K$-band data, $B$=1 indicates \Spitzer data.} 
 \label{par_table_continued}
 \end{table*}

\section{The Data}  

\subsection{BH Masses} 
We start with the compilation of 97 BH masses from \citet{S16}. Then, we add 3 galaxies (NGC1277, IC2560, Cygnus A) from the \citet{K13}'s compilation and other 5 galaxies (NGC1271, NGC1275, NGC1600, NGC3706, NGC5252) from \citet{VDB16}. Finally, we add the recent four BH mass estimates from \citet{Krajnovic18}. All BH masses are measured through either stellar dynamics, gasdynamics or astrophysical masers \citep{K13}, thus deriving from spatially resolved kinematics. We do not consider BHs with upper-limits on their masses or estimates from reverberation mapping or virial methods, since these methods need to be calibrated with the \mbhnosp-galaxy relations. This explains why our sample is a factor of 2 smaller than the largest BH masses compilations \citep{Beifiori12, VDB16}. We also discard BH masses  from papers where observational details are not provided (e.g. \citealt{Cappellari08}).\\
\indent Two strong matters of debate are given by the inclusion of DM halos (for stellar dynamics) and emission-line widths (for gasdynamics) in the analysis when estimating \mbhnosp. In both cases, the authors claim that neglecting these factors can yield underestimated \mbh values, even if in the first case not including the DM halo in the analysis only \textit{indirectly} affects \mbh through $M/L$ \citep{K13}. A DM halo is not always important (see e.g. Tab. 1 of \cite{Schulze11} or Tab. 3 of \cite{Rusli13}), especially if the BH sphere of influence is well resolved, while in the second case it is not guaranteed that the emission-line widths contribute significantly to the analysis, as they could simply be due to unresolved rotation (which is taken into account in the modelling) or turbolent motions. In this sample we find several galaxies having their \mbh estimated both with and without modeling a DM halo. When possible, we try to be conservative, keeping those estimates which take a DM halo into account (for stellar dynamics), or emission-line widths (for gasdynamics) in the analysis.
Details on our omissions are discussed in App.\tild\ref{Appendice}. This leaves us with a total of 83 galaxies (see Tab.\tild\ref{par_table}).

\subsection{Galaxy parameters} \label{gal_par}
\begin{figure*}
    	\includegraphics[width=\linewidth]{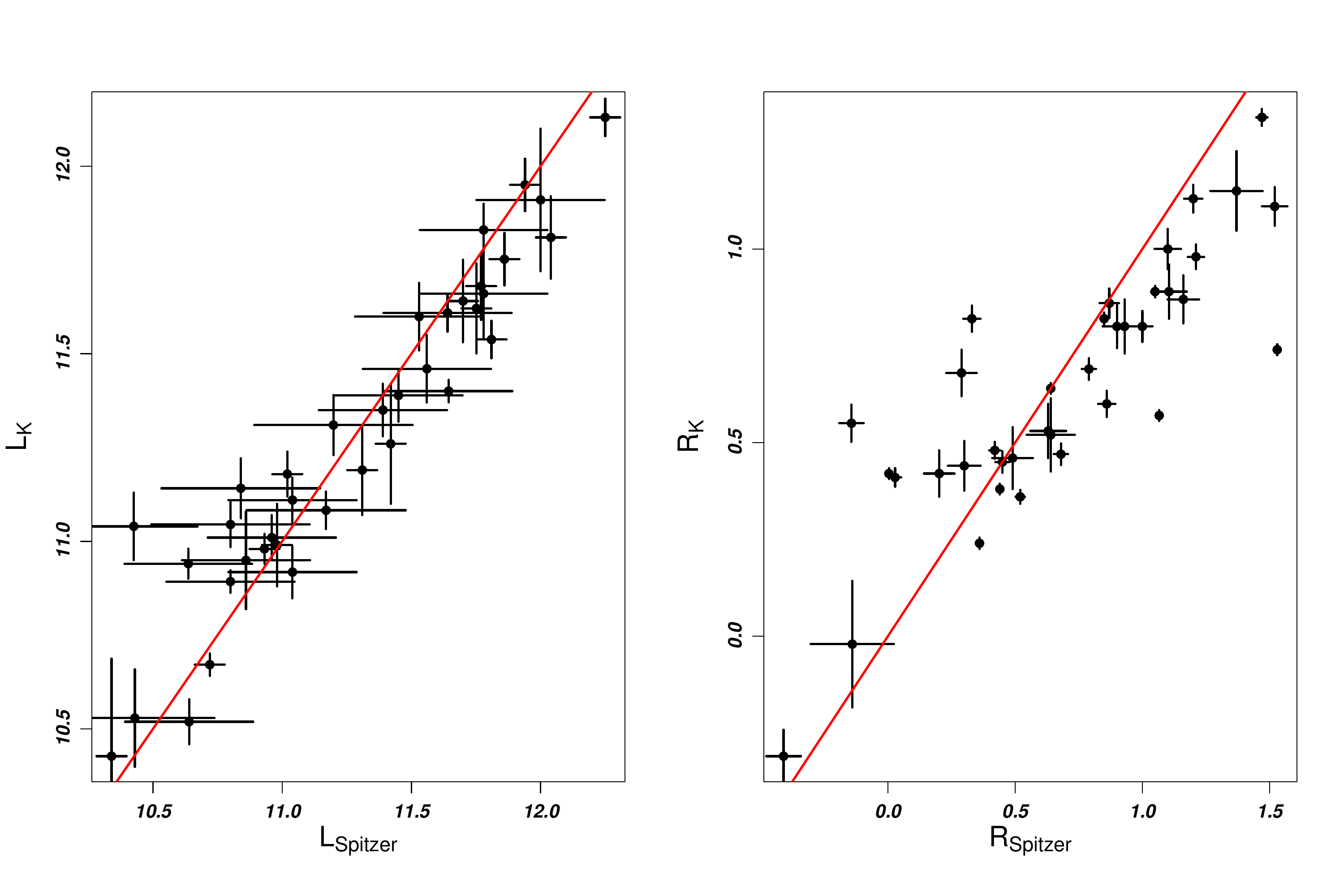}
	\caption{Comparison of the $K$-band and the \textit{Spitzer} photometries for the galaxies of our sample for which both measurements are avilable. The red lines are the 1:1 lines. The values are in good agreement with the \textit{Spitzer} radii being on average slightly larger, which can be expected since \textit{Spitzer} data are deeper. }
	\label{L_R}
\end{figure*}

Effective velocity dispersions are obtained from the same literature sources providing \mbhnosp\tild(\citealt{S16}, \citealt{K13} or \citealt{VDB16}). These are measured according to the equation \begin{equation}
\sigma = \left(\frac{\int_0^{R_e} \sqrt{\left(v(r)^2 + \sigma(r)^2\right)}I(r) dr} {\int_0^{R_e} I(r) dr}\right)^{\frac{1}{2}}
\label{sigma_comp}
\end{equation} 
where $v(r)$ and $\sigma(r)$ are the first two moments of the collision-less Boltzmann equation, $I(r)$ is the surface brightness profile and \re is the effective radius. We are convinced that the BH sphere of influence (hereafter SOI) should be included in the computation of \si since the BH itself is part of the system probed by the FP. Other authors (e.g. \citealt{McConnell13}) prefer to omit this region, in order to (try to) decouple the gravitational effects of the SMBH from the \si estimate.\\
\indent We take the 3.6 $\mu$m \Spitzer photometry from \citet{Savorgnan16a} (for the effective radii) and \citet{Savorgnan16b} (for the luminosities). In the first paper, the authors perform sophisticated decompositions, claiming not to underestimate the systematics involved in such analysis. We convert effective radii to physical units using distances of our sample. In order to alleviate the problem given by incomplete data, when \Spitzer photometry is not available we turn to the $K$-band photometry from 2MASS data performed by \citet{VDB16}. However, in this last work the focus is not on decompositions and hence such data can only be used for early-types. Moreover, these data are less deep than \textit{Spitzer}'s and the photometric analysis is performed in a much more simplified manner with respect to the work of \citet{Savorgnan16a}. Nevertheless, in Fig.\tild\ref{L_R} we compare \Spitzer and 2MASS data. For the luminosities, the agreement is fairly good ($\sim$0.12 dex), as can be expected given the low $K-3.6$ color index \citep{Sani11}, while things are slightly worse when dealing with radii ($\sim$0.20 dex), which can be explained by the different analyses and techniques used to derive the photometric variables \citep{VDB16, Savorgnan16a}.

\section{Linear regressions} \label{regr}
The first step of the analysis consists in fitting the scaling relations to our data. These relations have the form \begin{equation}
z = \alpha \left(x - \left<x\right>\right) + \gamma + \varepsilon
\label{linreg}
\end{equation}
for monovariate correlations and \begin{equation}
z = \alpha \left(x - \left<x\right>\right) + \beta \left(y - \left<y\right>\right) + \gamma + \varepsilon
\label{planereg}
\end{equation}
for bivariate correlations. In these equations, $\alpha$ and $\beta$ are the slopes, $\gamma$ is the zero-point and $\varepsilon$ is the intrinsic scatter around the dependent variable. This last parameter is of great importance since it represents the scatter not due to measurement errors, thus providing information about which variable(s) is (are) most closely connected to the central BH. Centering the independent variable reduces the covariance between the observables; moreover, the zero-point should be $\sim\left<z\right>$. Commonly used fitting routines are the Bayesian \texttt{linmixerr} and its multi-dimensional equivalent \texttt{mlinmixerr} \citep{Kelly07}. Here, we rely on the robust \texttt{lts\_linefit} and \texttt{lts\_planefit} algorithms \citep{Cappellari13}, which combine the Least Trimmed Squares algorithm from \citet{Rousseeuw06} with a residuals sum-of-squares minimization. These routines can automatically exclude the outliers from the fit, but we decided not to use this feature. For example, the largest galaxies (e.g. NGC4889, NGC1600, see Fig.\tild\ref{linregs}) are expected to be outliers of the \mbhnosp-\si relation because of the longer time needed by their SMBH to clear the bulge of gas, which results in abnormally large \mbhnosp's \citep{King15}. All the variables are logarithmic with units of measurement reported in Tab.\tild\ref{par_table}. In all fits, the dependent variable will be \mbhnosp.\\
\indent We consider the monovariate correlations between \mbh and \sinosp, \lk and \renosp. Then, we turn to bivariate correlations between \mbh and three possible pairs of galaxy parameters. The relation between \mbhnosp, \si and \re is of particular interest since it was proposed as the fundamental relation by \citet{Hopkins07b}. In the exhaustive study of \citet{S16}, this correlation is also detected, but with a stronger dependence on \si and a slightly weaker dependence on \renosp. 

\subsection{Regression results} \label{Sec.regression_results}
\begin{figure*}
{\includegraphics[scale=0.4]{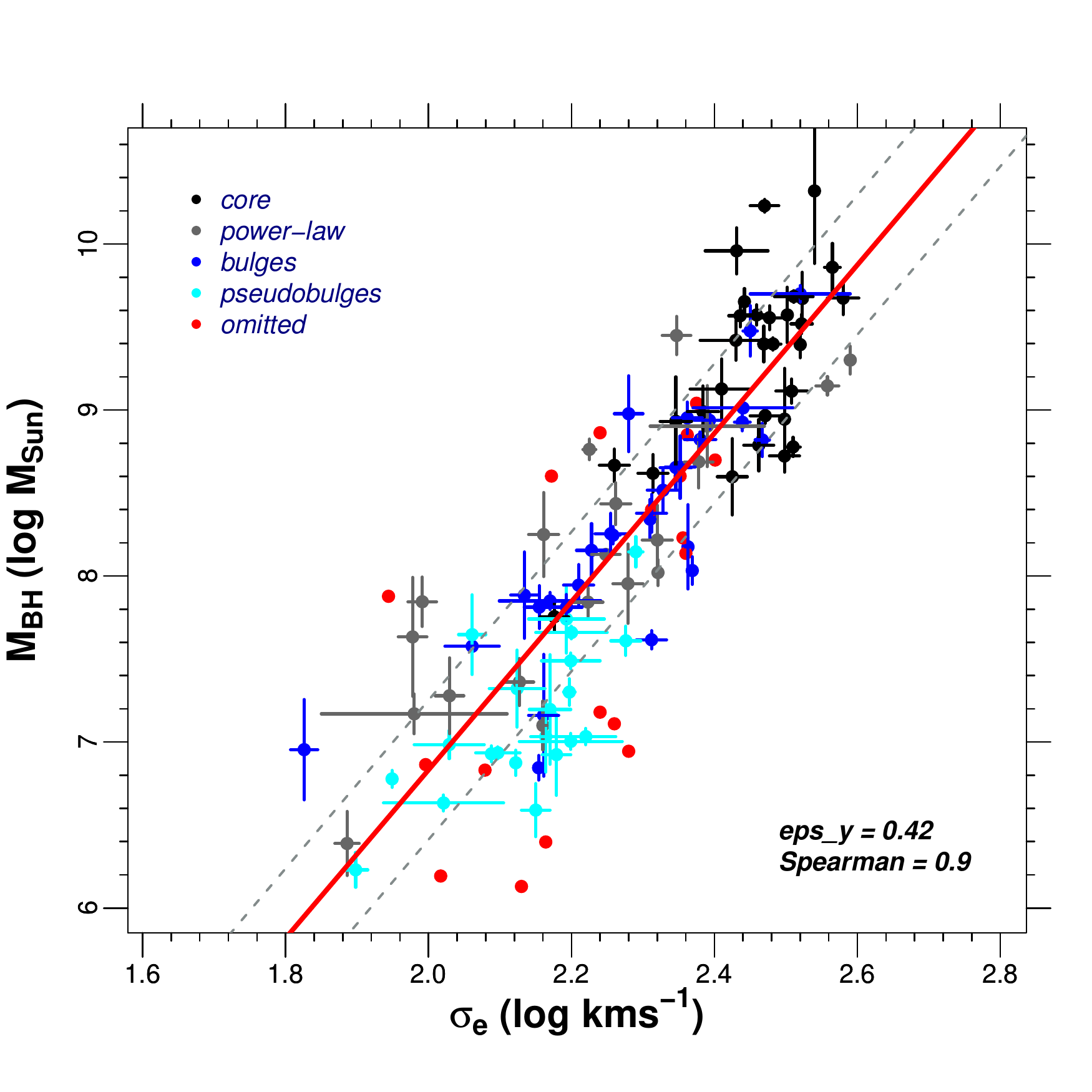}}

{\includegraphics[scale=0.4]{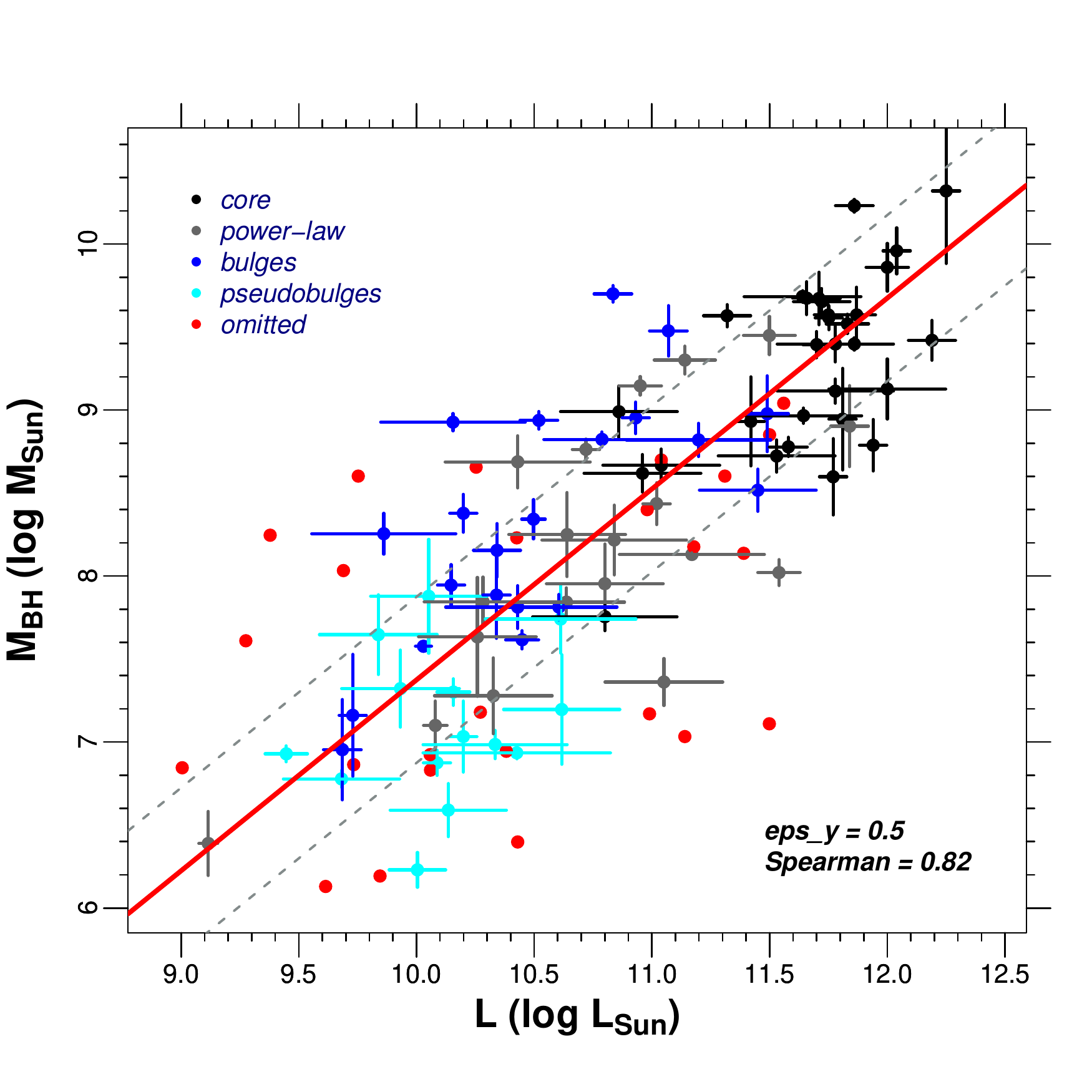}}
{\includegraphics[scale=0.4]{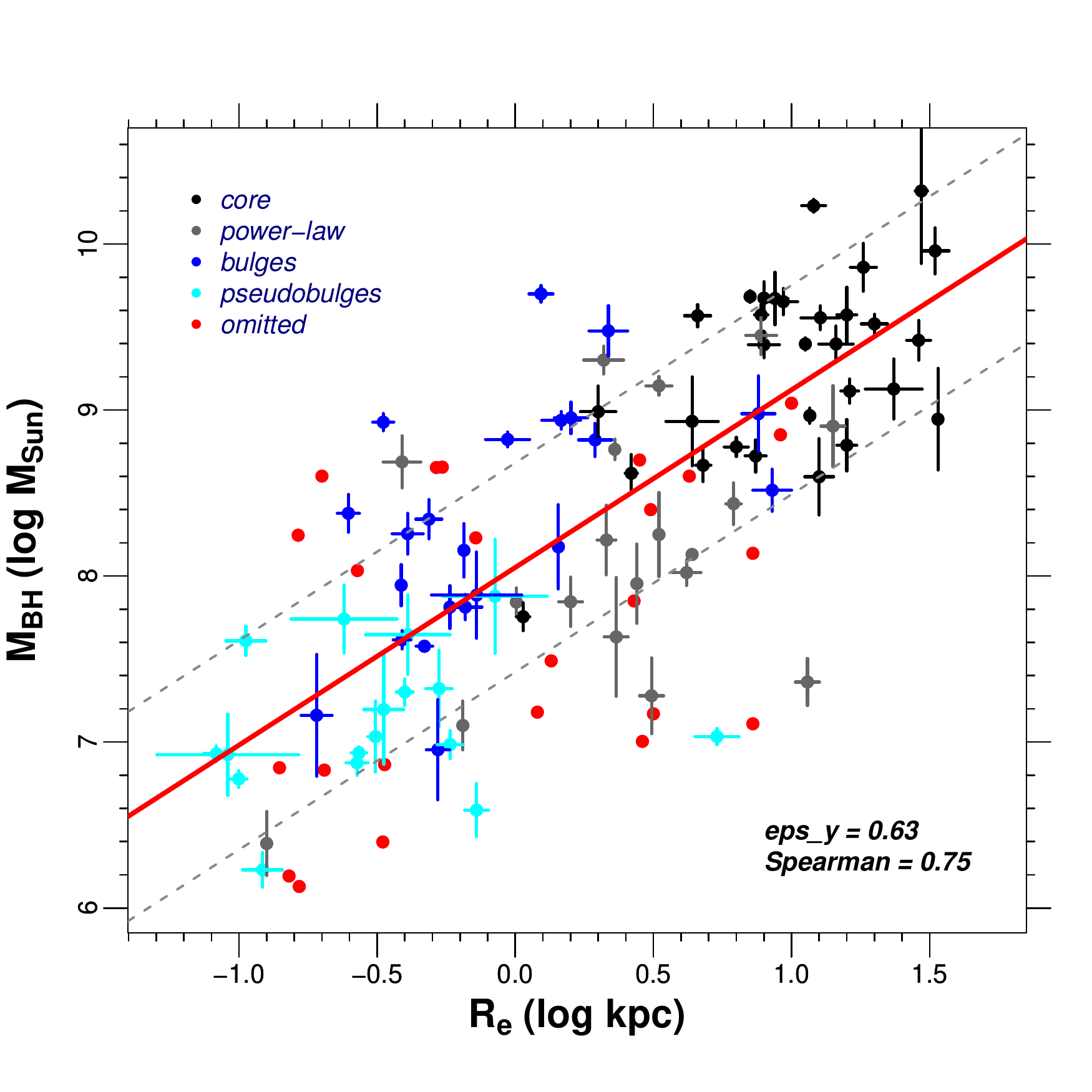}}
   \caption{Monovariate correlations between BH masses and galaxy parameters (\textit{upper row:} \sinosp, \textit{lower row, left:} $L$, \textit{lower row, right:} $R_e$). Galaxies are colored according to the $T$ flag defined in Tab. 1 of \citet{S16} (Col. 2 of Tab.\tild\ref{par_table_continued}). Red points are omitted from the regressions (see App.\tild\ref{Appendice}). The intrinsic scatter and the Spearman's coefficient are printed on the bottom-right of the plot. The dashed lines delimit the range given by the intrinsic scatter.}
	\label{linregs}
\end{figure*}

\begin{figure*} 
	{\includegraphics[width=.4\linewidth,height=.4\textheight]{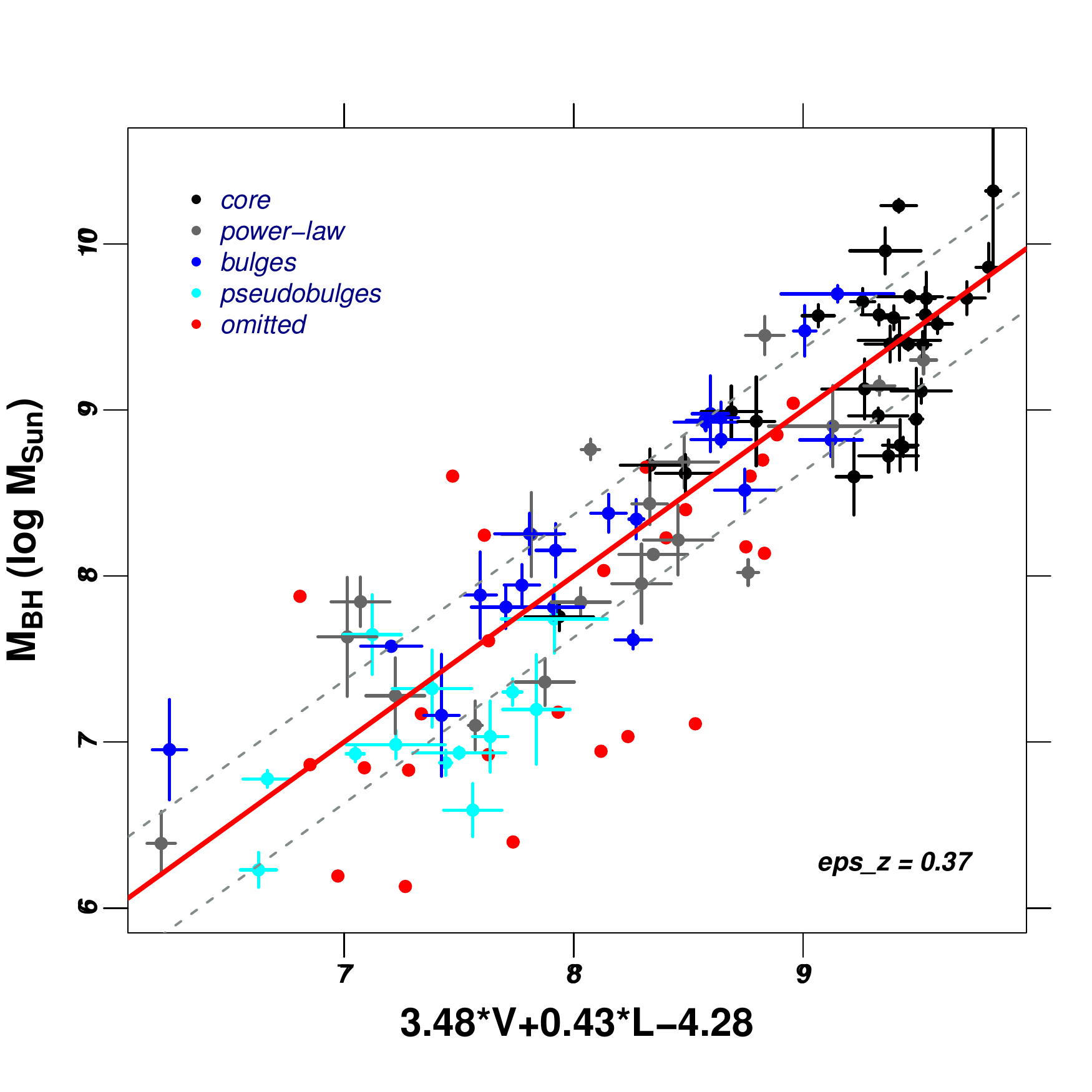}}
	{\includegraphics[width=.4\linewidth,height=.4\textheight]{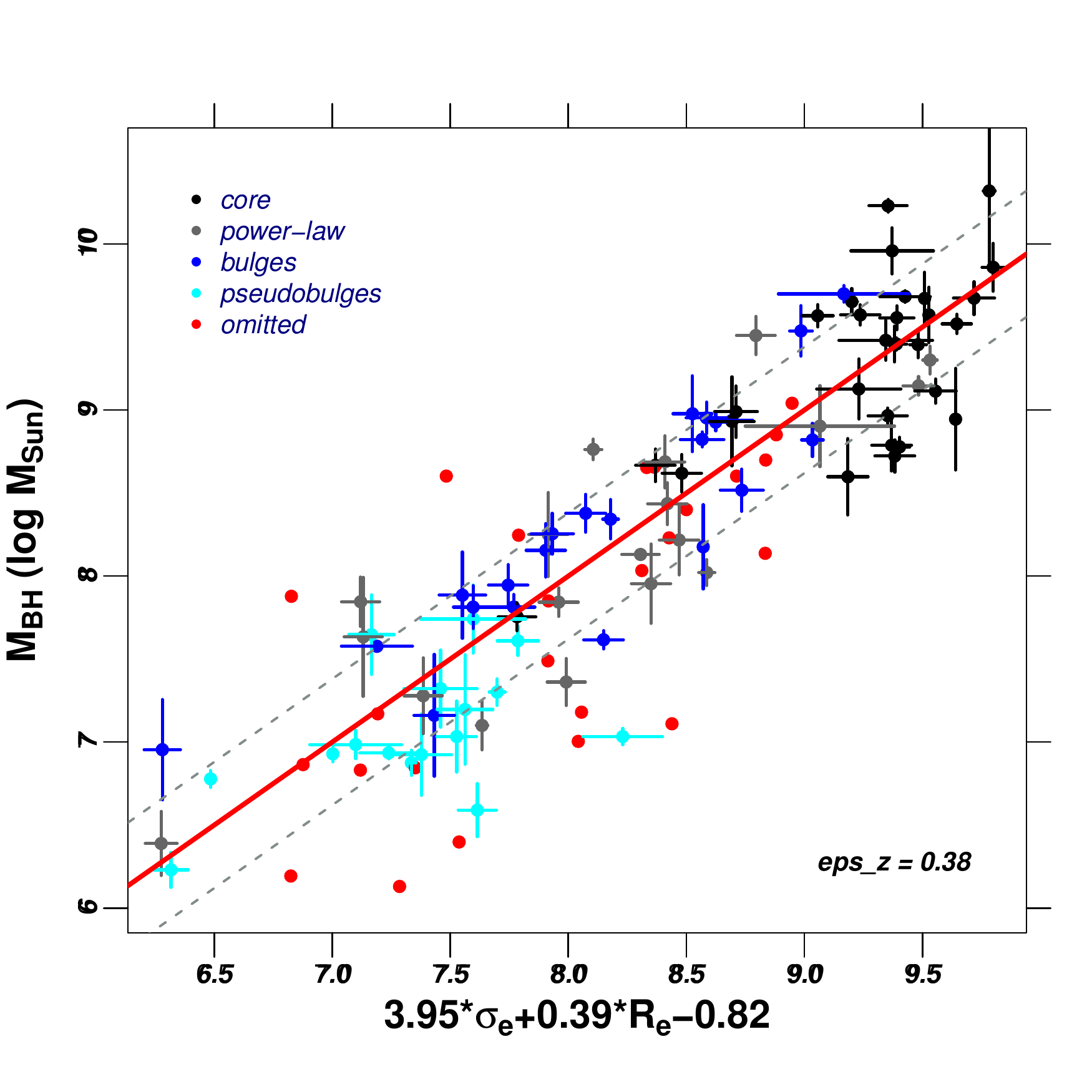}}
	
	\caption{Bivariate correlations between BH masses and galaxy parameters (\textit{left:} \mbhnosp-\sinosp-$L$, \textit{right:} \mbhnosp-\sinosp-$R_e$). Symbols and color coding are the same as in Fig.\tild\ref{linregs}.}
	\label{planeregs}
\end{figure*}

Regression results are reported in Tab.\tild\ref{regression_results}, and shown in Figs.\tild\ref{linregs} (monovariate) and \tild\ref{planeregs} (bivariate). The points interpreted as unreliable data (see App.\tild\ref{Appendice}) are plotted as red points and omitted from the regressions. Galaxies are divided into four subgroups (core ellipticals, power-law ellipticals, spirals with classical bulges and pseudobulges) using the $T$ flag defined in Tab. 1 of \citet{S16}. We notice that  \mbhnosp-\si has the lowest intrinsic scatter, whose value agrees with those found in similar studies \citep{Savorgnan16b, S16}. Using a sample of 45 early-types, \citet{K13} derive a relation with a scatter $<$0.3 $dex$. Instead, BHs correlate much more weakly with the bulge photometric parameters and, interestingly, the relations with luminosity and effective radius show the same slope. Since the intrinsic scatter embeds all factors not accountable with measurement errors, it appears that SMBHs are indeed more closely connected to \si than any other variable, confirming the conclusions of earlier works on this topic \citep{Gultekin09b, Beifiori12, VDB16}.\\
\indent At variance with their classical counterparts, pseudobulges do not seem to correlate with their central BHs \citep{Kormendy11}, except for a possible correlation with \sinosp. It is intriguing that in \citet{S16} this correlation is not detected (see their Table 11), even if their sample constitutes the basis of our own compilation\footnote{Note that we discard several galaxies which are taken into account in \citet{S16}'s analysis. See App.\tild\ref{Appendice}.}. It appears that the limited number of pseudobulges with reliable BH masses detections prevents us from reaching a definitive conclusion. Moreover, both classical and pseudobulges are not uniquely defined and several galaxies might host both components \citep{Erwin15}.\\ 
\indent It should be stressed that the large range spanned by BH 
masses can yield misleading results. In Fig. 11 of \citet{King15} it is shown how similar slopes but different normalizations for different galaxy subgroups can give anomalously high slopes. In Tab.\tild\ref{mbh_sigma_subgroups} we fit the \mbhnosp-\si relation to each of the four subgroups, showing that slopes are much closer to the value of 4 expected from momentum-driven theories, which do not constitute a serious threat for the bulge integrity. Since our BH mass estimates are biased towards higher values because of the limited resolution of current-days telescopes \citep{Bernardi07,Shankar16}, the slope of this relation could naturally increase with data from new generation telescopes, if the sample is very heterogeneous in masses. \\
\indent As shown in Tab.\tild\ref{regression_results}, combining \si with $L$ or \re does not significantly improve the intrinsic scatter of the \mbhnosp-\si relation. Nevertheless, neither $L$ nor \re have a slope consistent with zero, even at 3$\sigma$ limit. 
Our results are consistent with the findings of \citet{S16}\footnote{That authors use masses rather than luminosities. These are computed from mass-to-light ratios taken from different sources. See App. B of \citet{S16}.}. The dependence on \si is stronger than what originally found by \citet{Hopkins07b} and, interestingly, the \si coefficient is consistent with the value of 4 predicted by momentum-driven AGN feedback \citep{King03}.

\begin{table*}
\centering
\begin{tabular}{c c c c c c c c}
\hdr 
Variable(s) & Subgroup & $\alpha$ & $\beta$ & $\gamma$ & $\varepsilon$ & $\left<x\right>$ & $\left<y\right>$ \\ \hline
\si & \textit{All} & 5.07 $\pm$ 0.27 & - & 8.30 $\pm$ 0.05 & 0.42 $\pm$ 0.04 & 2.291 & - \\
& \textit{ClBul} & 4.48 $\pm$ 0.30 & - & 8.60 $\pm$ 0.05 & 0.38 $\pm$ 0.04 & 2.333 & - \\
    & \textit{Pseudo} & 3.50 $\pm$ 0.70 & - & 7.14 $\pm$ 0.07 & 0.27 $\pm$ 0.08 & 2.135 & - \\ 
   & & & & & & &\\
$L$ & \textit{All} & 1.12 $\pm$ 0.08 & - & 8.48 $\pm$ 0.06 & 0.53 $\pm$ 0.05 & 10.913 & - \\
& \textit{ClBul} & 1.00 $\pm$ 0.09 & - & 8.71 $\pm$ 0.06 & 0.47 $\pm$ 0.05 & 11.074 & - \\
    & \textit{Pseudo} & 0.49 $\pm$ 0.47 & - & 7.10 $\pm$ 0.12 & 0.43 $\pm$ 0.14 & 10.109 & - \\ 
   & & & & & & &\\
\re & \textit{All} & 1.07 $\pm$ 0.10 & - & 8.43 $\pm$ 0.07 & 0.63 $\pm$ 0.06 & 0.306 & - \\
& \textit{ClBul} & 0.91 $\pm$ 0.12 & - & 8.70 $\pm$ 0.07 & 0.60 $\pm$ 0.06 & 0.500 & - \\
    & \textit{Pseudo} & 0.15 $\pm$ 0.23 & - & 7.09 $\pm$ 0.10 & 0.39 $\pm$ 0.12 & -0.502 & - \\ 
 & & & & & & &\\
\sinosp-\re & \textit{All} & 3.95 $\pm$ 0.34 & 0.39 $\pm$ 0.09 & 8.43 $\pm$ 0.05 & 0.38 $\pm$ 0.04 & 2.301 & 0.310 \\
& \textit{ClBul} & 3.69 $\pm$ 0.34 & 0.32 $\pm$ 0.09 & 8.68 $\pm$ 0.05 & 0.34 $\pm$ 0.04 & 2.341 & 0.500 \\
    & \textit{Pseudo} & 2.80 $\pm$ 0.83 & -0.14 $\pm$ 0.19 & 7.05 $\pm$ 0.08 & 0.27 $\pm$ 0.11 & 2.120 & -0.529 \\ 
    & & & & & & &\\
\sinosp-$L$ & \textit{All} & 3.48 $\pm$ 0.43 & 0.43 $\pm$ 0.11 & 8.48 $\pm$ 0.05 & 0.37 $\pm$ 0.04 & 2.303 & 10.923 \\
& \textit{ClBul} & 3.20 $\pm$ 0.45 & 0.37 $\pm$ 0.11 & 8.69 $\pm$ 0.05 & 0.34 $\pm$ 0.04 & 2.341 & 11.074 \\
    & \textit{Pseudo} & 2.9 $\pm$ 1.3 & -0.08 $\pm$ 0.46 & 7.01 $\pm$ 0.10 & 0.30 $\pm$ 0.13 & 2.096 & 10.113 \\ \htw 
\end{tabular}
\caption{Regression results between \mbh and the galaxy parameters. \textit{Col. 1:} The independent variable(s). \textit{Col. 2:} The sample used in the regression, \textit{All} for the full sample, \textit{ClBul} of classical bulges and \textit{Pseudo} for pseudobulges. \textit{Cols. 3-4:} The slopes ($\alpha$ \& $\beta$). \textit{Col. 5:} The zero-point ($\gamma$). \textit{Col. 6:} The intrinsic scatter ($\varepsilon$). Cols. 7-8: The values of $\left<x\right>$ and $\left<y\right>$ (see eqs.\tild\ref{linreg}-\ref{planereg}).}
\label{regression_results}
\end{table*}

\begin{table*}
\centering
\begin{tabular}{c c c c c}
\hdr 
Subgroup & $\alpha$ & $\gamma$ & $\varepsilon$ & $\left<x\right>$ \\ \hline
Core & 4.32 $\pm$ 0.89 & 9.27 $\pm$ 0.07 & 0.38 $\pm$ 0.07 & 2.454 \\
Power & 3.65 $\pm$ 0.55 & 8.09 $\pm$ 0.10 & 0.41 $\pm$ 0.10 & 2.223 \\
ClBul & 4.25 $\pm$ 0.54 & 8.28 $\pm$ 0.07 & 0.34 $\pm$ 0.07 & 2.283 \\
Pseudo & 3.50 $\pm$ 0.70 & 7.14 $\pm$ 0.07 & 0.27 $\pm$ 0.08 & 2.135 \\ \htw 
\end{tabular}
\caption{Results of the fit of the \mbhnosp-\si relation to each subgroup defined in \citet{S16} (see also § \tild\ref{Sec.regression_results}. \textit{Col. 1:} The subgroup. \textit{Col. 2-3:} The slopes and the zero-points. \textit{Col. 4:} The intrinsic scatter. \textit{Col. 5:} The values of $\left<x\right>$ (see eq.\tild\ref{linreg}. All slopes are closer to the value of 4 predicted by momentum-driven theories \citep{King15} than that obtained from the full sample. Note the much larger uncertainty on the \textit{Pseudo} slope.}
\label{mbh_sigma_subgroups}
\end{table*}

\section{The BH hyperplane: A multivariate analytic approach} \label{modellino}
\indent This section describes a novel analytic approach which combines \mbh with the FP. The only other work published so far which deals with this issue is \citet{VDB16}, where the author shows that, when \si measurements are not available, the ratio $L_K/R_e$ should be used as a proxy of \mbhnosp. Here, we want to verify which relation gives the best prediction of the others, i.e. is able to reproduce their slopes and intrinsic scatters. Although we know that the FP does not improve if additional parameters are added \citep{Djorgovski87}, in order to find the fundamental BH-hosts relation, this plane must be taken into account since it provides the most general description of a host bulge. Moreover, the BH itself could contribute to the tilt in the FP (see \citealt{VDB16}). First, we model the FP with a trivariate Gaussian following \citet{Bernardi03b}. Then, we show how to use this description to find the fundamental relation. Since pseudobulges do not seem to follow the scaling relations, we have preferred to omit them from the sample. Results including these systems are reported in App.\tild\ref{Appendice2}, where we show that our conclusions do not differ significantly.

\subsection{A four-dimensional regression} \label{data_analysis}
\begin{figure} 
	\centerline{\includegraphics[width=\linewidth]{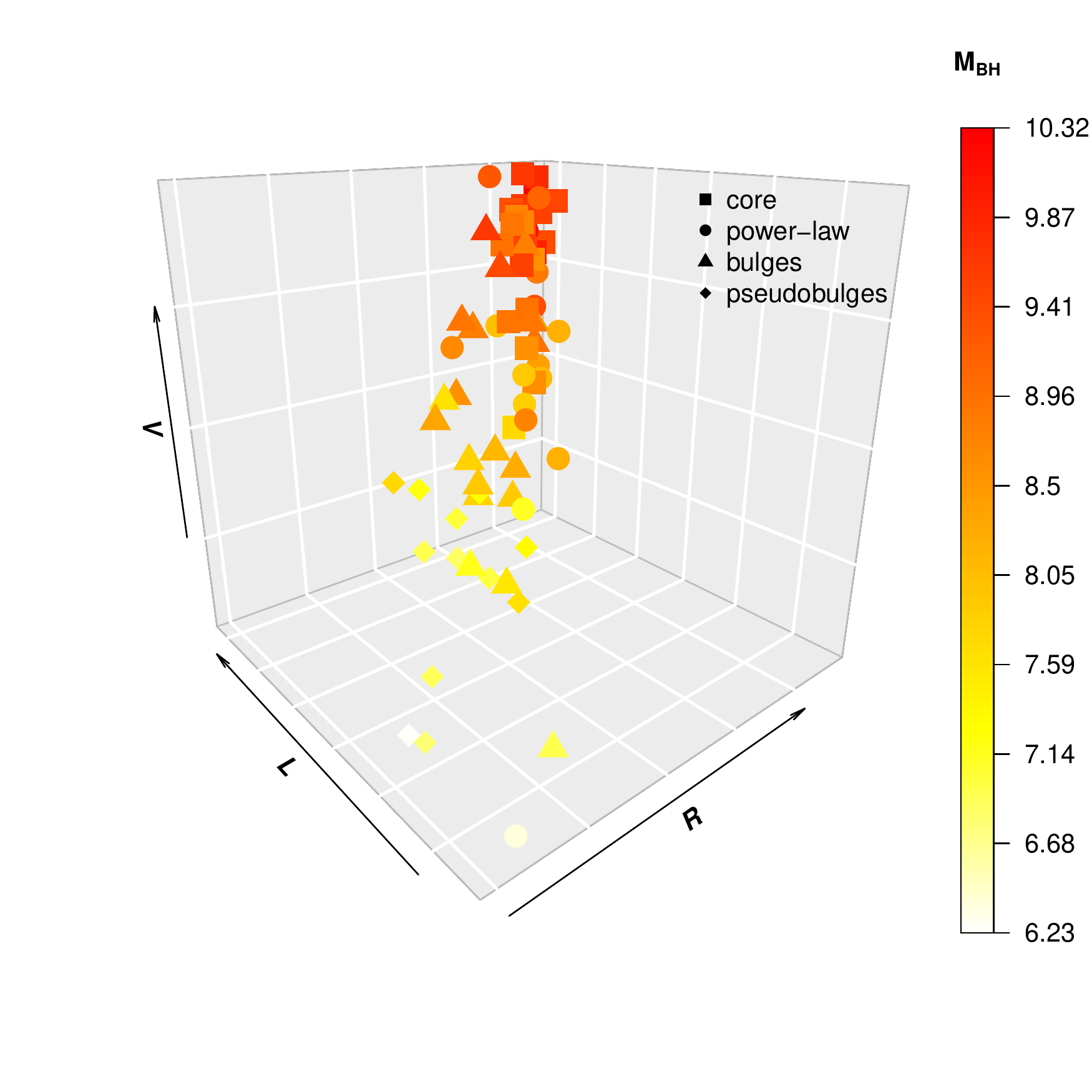}}
	\caption{The 3D representation of the putative ``BH hyperplane''. Different symbols denote different galaxy subsets, as shown in the legend. Points are colored according to their \mbh values.}
	\label{Fig.hyperplane}
\end{figure}

\indent We start writing down the most general relation between \mbh and the host galaxy\footnote{To avoid confusion with the standard deviation of the gaussians, we will use V to label the velocity dispersion in the remainder of the paper.}:
\begin{equation}
\mbhmath = \alpha\,L + B\,R_e + C\,V + g_0\,\Sigma
\label{hyperplane}
\end{equation}
which is shown in Fig.\tild\ref{Fig.hyperplane} for our sample. All the variables of appearing in this equation (\mbhnosp, $L$, $V$ and $R_e$) are logarithmic. Here $\Sigma$ is the dispersion of this relation and $g_0$ a random gaussian number with zero mean, so that the product of $\Sigma$ and $g_0$ represents the intrinsic scatter of our relation.\footnote{Note that there should also be a dependence on redshift, but since for our sample $D_{max}$ $\sim$ 250 $Mpc$ (Cygnus A), this term is negligible.}\\
\indent The first thing to do is fit eq.\tild\ref{hyperplane} to our data, in order to check if we are able to constrain the three slopes of the hyperplane and its intrinsic scatter. This is needed not only to see how the coefficients compare to those obtained for lower-dimensionality relations but also to quantify the effect of the FP in establishing such correlations. The regression has been performed by extending the fitting routine used in the previous section \citep{Cappellari13} to the four-dimensional case. We obtain:
\[A = -0.12 \pm 0.33;\]
\[B = 0.56 \pm 0.33;\]
\[C = 4.18 \pm 0.48;\]
\[\Sigma = 0.35 \pm 0.04;\]
\[N = 8.34 \pm 0.05,\] 
where $N$ is the zero-point of the regression. The errors are only slightly larger than those obtained with planar regressions, which is reassuring because the dimensionality of the problem combined with the low number of points could have yielded abnormally large errors, or even prevented the convergence of the algorithm.\\
\indent We see that the intrinsic scatter of this relation is comparable with those of the other regressions where \si is involved (see col. 6 of Tab.\tild\ref{regression_results}), resulting slightly lower than the BHFPs with \si and consistent with the value found for \mbhnosp-\si within 1.5$\sigma$. The comparison between the scatters is made by evaluating the quantity
\begin{equation}
    \frac{|\varepsilon_1 - \varepsilon_2|}{\sqrt{\sigma_{\varepsilon_1}^2 + \sigma_{\varepsilon_2}^2}}.
    \label{discrepanza}
\end{equation}
\noindent Once again, it appears that \si alone is a very good predictor of \mbhnosp. Interestingly, the introduction of $L$ does not significantly alter the slopes of the $\mbhmath-\simath-\remath$, indeed the $L$-slope is the only one consistent with zero within 1$\sigma$. 

\subsection{Modelling the BH hyperplane} \label{ss42}

\indent We now investigate the effects of the FP on the relations between BH and bulge structural parameters. The three variables defining the FP are strongly pairwise correlated \citep{Bernardi03b}. Due to the smallness of our sample, computing covariances and correlations between these three observables may be cumbersome. Thus, we turn to the much more robust analysis of \citet{Bernardi03b}. 

\begin{figure*}
\centering
     \includegraphics[width=.9\paperwidth,height=.5\paperheight]{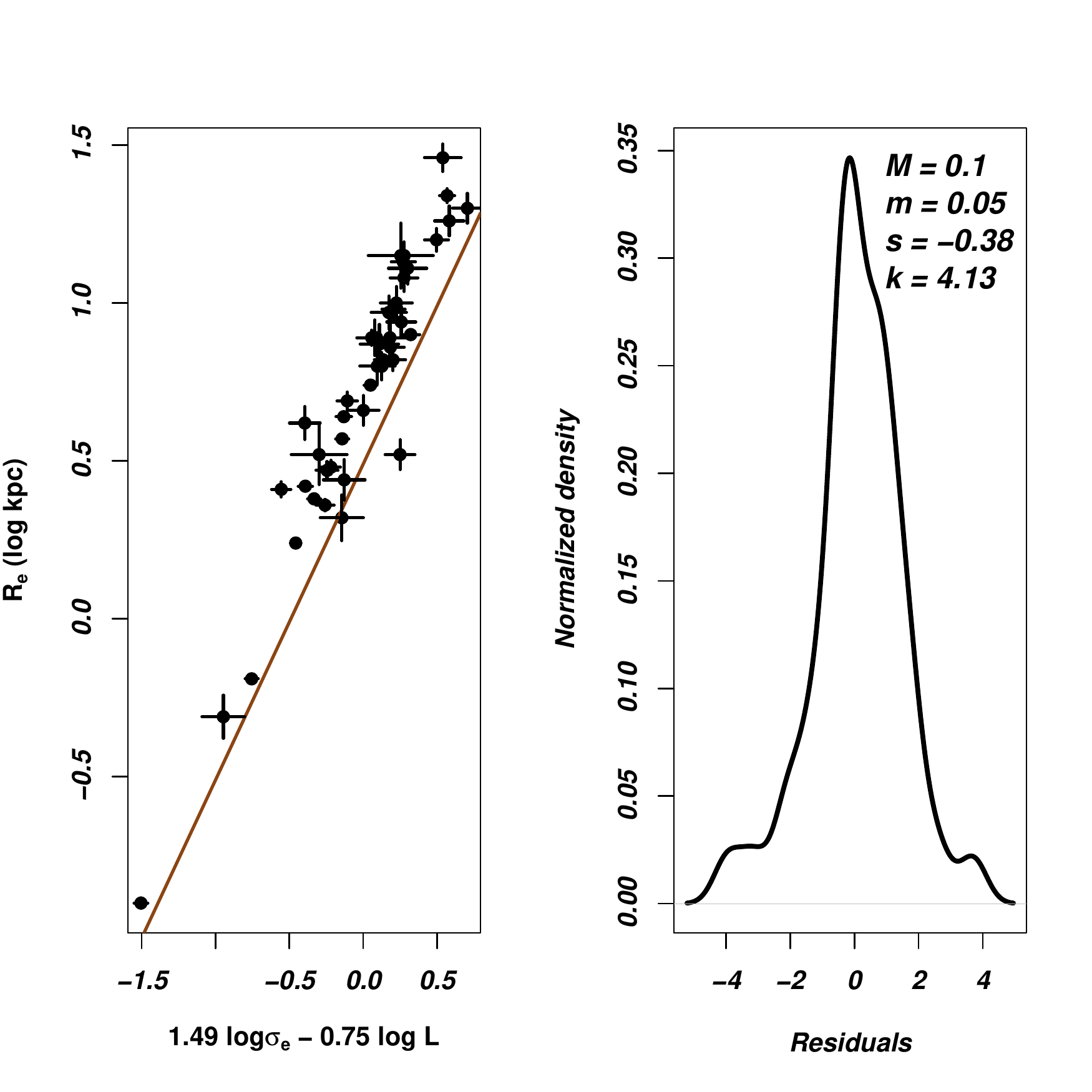}
		\caption{\textit{Left:} Effective radius $R_e$ as a function of the FP relation found by \citet{Bernardi03c}. Our galaxies follow that FP, whose equation is reported on the x-axis, but are larger than expected. \textit{Right:} Distribution of the residuals (normalized to unity) with respect to to the new best-fit line obtained by minimizing with respect to the normalization only. $M, m, s, k$ are mean, median, skewness and kurtosis, respectively. Both quantities on the axes are unit-less}.
	\label{debias}
    \end{figure*}

\noindent In that work, the authors studied a sample of $\sim$9000 early-type galaxies from the SDSS finding that the luminosity distribution is well modeled by a gaussian (see also \citealt{Saglia01}), and also the distributions of both $V$ and $R_e$ around the mean (at fixed $L$) are of gaussian shape. This means that \begin{equation}
\phi(L, R_e, V) = \psi(R_e, V | L)\phi(L)
\label{distribution}
\end{equation} 
where $\psi(R_e, V | L)$ and $\phi(L)$ are a bi- and a monovariate gaussian, respectively, and $\phi(L, R_e, V)$ is the joint distribution, which is well modeled by a trivariate gaussian. In practice, we draw $L$ from a gaussian distribution with mean $\left<L\right>$ and variance $\sigma^2_L$ then $R_e$ from a gaussian with mean $\left<R_e|L\right>$ and variance $\sigma^2_{R_e|L}$ and finally the velocity dispersion taking into account both correlations with $L$ and $R_e$. Labeling the correlation coefficients between two variables with $\rho$, we obtain (see Appendix A of \citealt{Bernardi03b})
\begin{subequations} \label{multivariate_correlation}
	\begin{align}
	L &= g_1\sigma_L     \\
	R_e &= \frac{L}{\sigma_L}\sigma_{R_e}\rho_{R_eL} + g_1\sigma_{R_e}\sqrt{1 - \rho^2_{R_eL}} \label{LR} \\
	V &= \frac{L}{\sigma_L} \xi_{LV} + \frac{R_e}{\sigma_{R_e}} \xi_{R_eV} + g_2\sigma_{V|R_eL},
	\end{align}
\end{subequations}

\noindent where the $g$'s are Gaussian random numbers with zero mean and unit variance,
\newpage
\begin{subequations} \label{help}
	\begin{align}
	&\xi_{LV} = \sigma_V \frac{\rho_{LV} - \rho_{R_eL}\rho_{R_eV}}{1 - \rho^2_{LR_e}} \\
	&\xi_{R_eV} = \sigma_V \frac{\rho_{R_eV} - \rho_{R_eL}\rho_{LV}}{1 - \rho^2_{LR_e}} \\
	&\sigma_{V|LR_e} = \sigma_V\sqrt{\frac{1 - \rho^2_{LR_e} - \rho^2_{LV} - \rho^2_{VR_e} + 2\rho_{R_eL}\rho_{LV}\rho_{R_eV}}{1 - \rho^2_{LR_e}}}
	\end{align}
\end{subequations}
\noindent and

\begin{equation}
   \mathscr{C} = 
\begin{pmatrix}
\sigma_L & \rho_{R_eL} \sigma_{R_e} \sigma_L & \rho_{LV} \sigma_L \sigma_V \\
\rho_{R_eL} \sigma_{R_e} \sigma_L & \sigma_{R_e} & \rho_{R_eV} \sigma_{R_e} \sigma_V \\
\rho_{LV} \sigma_L \sigma_V & \rho_{R_eV} \sigma_{R_e} \sigma_V & \sigma_V
\end{pmatrix} 
\end{equation}

\noindent is the covariance matrix. The parameters of this matrix should be estimated using a maximum likelihood analysis: since we expect our data to be distributed following a trivariate normal distribution
\begin{equation}
\mathscr{L} = \mathscr{N}^{3D}(\left\{\bm{L,V,R_e}\right\},\mathscr{C}),
\label{trivariate}
\end{equation}
where $\mathscr{C}$ is the covariance matrix, we can use this function to estimate the six parameters which define $\mathscr{C}$ needed to determine our best-fit function. However, since our sample is not very large, we speculate that these six parameters might be so well constrained. In order to increase the robustness of the analysis, we can take the covariance matrix derived from the sample of \citet{Bernardi03b}. In fact, that covariance matrix describes the properties of a generic sample of early-types. If we consider only the early-types in our sample, then we should, in principle, deal with a (biased) subset of that sample. The problem is that our photometric data are either at 3.6$\mu m$ or in the $K$-band, but such a covariance matrix is not available for that bands. To alleviate the problem given by the fact that the SDSS observations are carried out in the optical, we take the covariance matrix from \citet{Bernardi03b} in the $z$-band (see Tab.\tild\ref{coeffs}), thus assuming that both variances and correlations do not change significantly. \\
\indent As a test to ensure that our FP is consistent with that derived in \citet{Bernardi03c}, we plot in the left panel of Fig.\tild\ref{debias} the early-types of our sample with their best-fit line. Our galaxies seem to follow that FP but are, on average, larger than expected. This is a selection effect generated by the need of large galaxies to resolve the BH SOIs, otherwise the required resolution is beyond current-day facilities \citep{Bernardi07,Shankar16}. In order to remove the bias, we estimated the normalization through a one-dimensional regression fixing the slopes to the FP of \citet{Bernardi03c} and then computed the residuals with respect to the new best-fit line. 

\begin{table}
	\centering
	\begin{tabular}{|c c|}
		\hdr 
		Coefficient & Value \\ \hline
		$\sigma_V$ & 0.17 \\
		$\sigma_L$ & 0.69 \\
		$\sigma_{R_e}$ & 0.64 \\
		$\rho_{VL}$ & 0.78 \\
		$\rho_{VR_e}$ & 0.54 \\
		$\rho_{R_eL}$ & 0.88 \\ \htw 
	\end{tabular}
	\caption{The covariance matrix describing the FP of the sample of \citet{Bernardi03b}. This is derived from $z$-band observations.}
	\label{coeffs}
\end{table}

The residuals distribution is plotted in the right panel of Fig.\tild\ref{debias}. The symmetry of the residuals plot provides validation of our early-types being a (biased high) subset of the 9000 early-types of \citet{Bernardi03a}.\\
\indent The next step consists in substituting eqs.\tild\ref{multivariate_correlation} into eq.\tild\ref{hyperplane}. To this extent, we have developed a \textit{Mathematica} code which combines eq.\tild\ref{hyperplane} with eqs.\tild\ref{help}. We obtain: 
\begin{equation}
M_{BH} = \alpha_L L + g_0\varepsilon_L
\label{mbhlum}
\end{equation}
where
\begin{subequations} \label{L}
	\begin{align}
	&\alpha_L = A + \frac{B\sigma_{R_e}\rho_{R_eL} + C\sigma_V\rho_{VL}}{\sigma_L}, \label{sl_lum_analytic}\\ 
	&\varepsilon_L = \sqrt{\Sigma^2 + B^2\sigma^2_{R_e} (1-\rho^2_{R_eL}) + C^2\sigma^2_V (1-\rho^2_{VL}) + \Gamma_L}, \label{scatt_lum_analytic}\\ 
 &\Gamma_L =  2BC\sigma_{R_e}\sigma_V (\rho_{R_eV} - \rho_{LV}\rho_{R_eL}).
	\end{align}
\end{subequations}

\noindent We have thus put together all the terms multiplying $L$ and all those multiplying the casual coefficients by adding them in quadrature, so that $\alpha_L$ is the slope of the relation and $\varepsilon_L$ its intrinsic scatter. In practice, we are deriving the analytic expression $\left(\mbhmath | V,\,R_e\right)$ relation, i.e. we are projecting the hyperplanar relation of the previous paragraph on the \mbhnosp-$L$ relation. Because of the symmetry of the trivariate distribution, we can write analogous expressions to eq.\tild\ref{mbhlum} by simply interchanging variables and coefficients. If we drew, say, $L$ and $V$ from $R_e$, then: \begin{equation}
M_{BH} = \alpha_{R_e}R_e + g_0\varepsilon_{R_e}
\label{mbhre}
\end{equation} 
where 
\begin{subequations} \label{R}
	\begin{align}
	&\alpha_{R_e} = B + \frac{A\sigma_L\rho_{R_eL} + C\sigma_V\rho_{VR_e}}{\sigma_{R_e}}, \label{sl_re_analytic}\\
	&\varepsilon_{R_e} = \sqrt{\Sigma^2 + A^2\sigma^2_L (1-\rho^2_{R_eL}) + C^2\sigma^2_V (1-\rho^2_{VR_e}) + \Gamma_{R_e}}, \label{scatt_re_intrinsic}\\
& \Gamma_{R_e} = 2AC\sigma_L\sigma_V (\rho_{LV} - \rho_{R_eV}\rho_{R_eL}) ,
	\end{align}
\end{subequations}

\noindent while starting from $V$ we would obtain \begin{equation}
\mbhmath = \alpha_VV + g_0\varepsilon_V
\label{mbhsig}
\end{equation} 
where 
\begin{subequations} \label{V}
	\begin{align}
	&\alpha_V = C + \frac{A\sigma_L\rho_{VL} + B\sigma_R\rho_{VR_e}}{\sigma_V}, \label{sl_sig_analytic}\\
	&\varepsilon_V = \sqrt{\Sigma^2 + A^2\sigma^2_L (1-\rho^2_{R_eV}) + B^2\sigma^2_{R_e} (1-\rho^2_{VR_e}) + \Gamma_V}, \label{scatt_sig_intrinsic}\\
&\Gamma_V = 2AB\sigma_L\sigma_{R_e} (\rho_{R_eL} - \rho_{R_eV}\rho_{VL}).
	\end{align}
\end{subequations}

\noindent Eqs.\tild\ref{L}, \ref{R} and \ref{V} are those we are going to use to find the slopes of and the intrinsic scatters starting from the intrinsic relation\tild\ref{hyperplane} and taking the FP into account. 

\subsection{The fundamental BH-hosts relation}
Using eqs.\tild\ref{sl_lum_analytic}, \ref{sl_re_analytic} and \ref{sl_sig_analytic} we can compute the expected slopes for the three monovariate relations using the values of $A$, $B$ and $C$ from the hyperplanar regression, while eqs.\tild\ref{scatt_lum_analytic}, \ref{scatt_re_intrinsic} and \ref{scatt_sig_intrinsic} give us the three intrinsic scatters. However, considering the approximations we made (see par.\tild\ref{ss42}), we preferred to be conservative trying to simplify this approach by \textit{assuming one relation to be fundamental} and see how it performs in predicting the others.\\
For example, let us assume \mbhnosp-$V$ to be the fundamental relation. If $V$ is the sole variable of importance in deriving \mbhnosp, then we should have $A=B=0$ in eq.\tild\ref{hyperplane} and $C$ \& $\Sigma$ equal to slope and intrinsic scatter of the \mbhnosp-$V$ (Tab.\tild\ref{regression_results}). We must set $A=B=0$ not only in eq.\tild\ref{hyperplane}, but also in eqs.\tild\ref{R} and eqs.\tild\ref{L}. Then, by using for $C$ and $\Sigma$ the values obtained through the linear regression we can check how $V$ predicts slopes and intrinsic scatters of \mbhnosp-$L$ (using eqs.\tild\ref{L}) and \mbhnosp-$R_e$ (using eqs.\tild\ref{R}).\\ 
\indent This procedure must then be repeated assuming, in turn, the other two relations to be fundamental. Besides these three photometric quantities appearing in our sample, the results of above suggest considering linear combinations of $V$ and $R_e$ too, i.e. \[W = aV + bR_e\] where $a$ and $b$ are integers which give $W$ a particular meaning. The three combinations examined here are: \begin{itemize}
	\item $a$ = 2 and $b$ = 1, which make $W$ the mass predicted by the virial theorem ($M_{vir}$);
	\item $a$ = 4 and $b$ = 0.4 ($M_{Hop}$), which is the relation proposed by \citet{Hopkins07b} as the fundamental relation with the coefficients derived in Tab.\tild\ref{regression_results};
	\item $a$ = 4 and $b$ = 1, which makes $W$ the gravitational energy $U_{grav}$ of a singular isothermal sphere (SIS). 
\end{itemize}
These three quantities can be used in eq.\tild\ref{hyperplane} just as $L$, $V$ and $R_e$. In particular, \mvir is expected to be a good proxy of \mbul \citep{Cappellari06}. The results of the regressions (using the whole sample) linking \mbh to these three new variables (eq.\tild\ref{linreg}) are reported in Tab.\tild\ref{regression_results_2}. Apart from the \mbhnosp-$M_{Hop}$, no relation is better than the \mbhnosp-\si in terms of intrinsic scatter.

\begin{table*}   
	\centering
	\begin{tabular}{c c c c c c}
		\hline \hline
		Variable(s) & Subgroup & $\alpha$ & $\beta$ & $\varepsilon$ & $\left<x\right>$ \\ \hline
		\Hop & \textit{All} & 0.99 $\pm$ 0.05 & 8.43 $\pm$ 0.04 & 0.37 $\pm$ 0.04 & 9.332\\
		& \textit{ClBul} & 0.88 $\pm$ 0.05 & 8.85 $\pm$ 0.04 & 0.35 $\pm$ 0.04 & 9.504 \\
		& \textit{ET}  & 0.92 $\pm$ 0.06 & 8.85 $\pm$ 0.04 & 0.38 $\pm$ 0.04 & 9.689\\
		& & & & \\
		\mvir & \textit{All} & 0.88 $\pm$ 0.05 & 8.44 $\pm$ 0.05 & 0.46 $\pm$ 0.05 & 4.813 \\
		& \textit{ClBul} & 0.80 $\pm$ 0.05 & 8.83 $\pm$ 0.04 & 0.45 $\pm$ 0.05 & 4.903\\
		& \textit{ET}  & 0.93 $\pm$ 0.06 & 8.84 $\pm$ 0.04 & 0.44 $\pm$ 0.04 & 5.012 \\
		& & & & \\
		\mgrav & \textit{All} & 0.69 $\pm$ 0.04 & 8.43 $\pm$ 0.05 & 0.41 $\pm$ 0.04 & 9.513\\ 
		& \textit{ClBul} & 0.63 $\pm$ 0.05 & 8.83 $\pm$ 0.04 & 0.39 $\pm$ 0.04 & 9.679\\
		& \textit{ET}  & 0.69 $\pm$ 0.04 & 8.84 $\pm$ 0.04 & 0.40 $\pm$ 0.04 & 9.824 \\ \htw
	\end{tabular}
\caption{Regression results between \mbh and quantities obtained from linear combinations of \si and \renosp. \textit{Col.1}: the independent variable. \textit{Col.2}: the subgroup. \textit{Col.3}: the slope. \textit{Col.4}: the zero-point. \textit{Col.5}: the intrinsic scatter. \textit{Col.6}: The value of $\left<x\right>$ (eq.\tild\ref{linreg}). The subgroup \textit{ET} embeds galaxies flagged with $A = 0$ (core-galaxies) or $A = 1$ (power-law ellipticals, Tab.\tild\ref{par_table}).}
\label{regression_results_2}
\end{table*}

In order to assess the goodness of the predictions, a $\chi^2$ defined as \begin{equation}
\chi^2 = \sum_j \left[\left(\frac{\alpha_{obs} - \alpha_{mod}}{\sigma_{\alpha,obs}}\right)^2 + \left(\frac{\varepsilon_{obs} - \varepsilon_{mod}}{\sigma_{\varepsilon,obs}}\right)^2\right] 
\label{chisq}
\end{equation}  
has been used to compare the results predicted by a relation with those obtained from the fits. The subscripts $obs$ and $mod$ refer to the parameters derived from the regression and those computed with our model, respectively. The variances of the three new variables of Tab.\tild\ref{regression_results_2} are linked to those of $V$ and $R_e$ by the notorious formula 
\begin{equation}
    \sigma_W^2 = a^2\sigma_V^2 + b^2\sigma_{R_e}^2 + 2ab\sigma_V\sigma_R\rho_{VR_e}.
\end{equation}

\indent The results for the sample of early-type galaxies are shown Tab.\tild\ref{modellino_mine}. $V$ predicts the other relations better than its linear combinations with $R_e$ or the other two monovariate correlations. It is intriguing that, even though $R_e$ alone completely fails in predicting the coefficients of the other regressions, increasing its exponent from 0.4 ($M_{Hop}$) to 1 ($U_{grav}$) does not change the $\chi^2$ significantly, showing that $V$ is the predominant variable. \\
We now repeat the analysis considering \textit{all} classical bulges, i.e. also those coming from decompositions of spirals. Although we do not have a covariance matrix that describes such a sample, we can still take the covariance matrix of \citet{Bernardi03b}, since classical bulges behave in the same way as early-types. We caution that in this case all the errors coming from the decompositions are additional sources of uncertainty for the results that follow, even though we will show in the next paragraph that what really makes the difference for the results is the covariance matrix.\\
\indent Taking again the values of Tab.\tild\ref{coeffs} to build the covariance matrix, we obtain the results reported in Tab.\tild\ref{modellino_mine_clbul}.
We see that the only difference with respect to the previous paragraph is that here the results of $M_{Hop}$ and $U_{grav}$ are interchanged. But even in this case it is the velocity dispersion to yield the best predictions of the other relations. 

\begin{table}  
	\centering
	\setlength{\tabcolsep}{3pt}
	\begin{tabular}{|c|c c c c c c c c|}
		\hdr
	 Fund. / Obs. & & $L$ & $R_e$ & $V$ & $M_{Hop}$ & $M_{Vir}$ & $U_{Grav}$ & $\chi^2$ \\ \hline    
		$L$ & $\alpha$ & \textbf{1.13} & 1.38 & 2.66 & 0.68 & 0.90 & 0.59 & 34.1 \\
		& $\varepsilon$ & \textbf{0.44} & 0.47 & 0.50 & 0.47 & 0.45 & 0.46 &\\
		& & & & & & & &\\
		$R_e$ & $\alpha$ & 0.73 & \textbf{1.14} & 1.35 & 0.43 & 0.64 & 0.40 & 150.5\\
		& $\varepsilon$ & 0.61 & \textbf{0.59} & 0.64 & 0.62 & 0.60 & 0.61 &\\
		& & & & & & & &\\
		$V$ & $\alpha$ & 1.11 & 1.08 & \textbf{4.32} & 0.93 & 1.01 & 0.73 & 1.4\\
		& $\varepsilon$ & 0.50 & 0.57 & \textbf{0.41} & 0.41 & 0.47 & 0.44 &\\
		& & & & & & & &\\
		$M_{Hop}$ & $\alpha$ & 1.24 & 1.48 & 4.07 & \textbf{0.92} & 1.07 & 0.74 & 7.4\\
		& $\varepsilon$ & 0.43 & 0.48 & 0.39 & \textbf{0.38} & 0.41 & 0.39 &\\
		& & & & & & & &\\
		$M_{Vir}$ & $\alpha$ & 1.10 & 1.49 & 2.96 & 0.71 & \textbf{0.93} & 0.62 & 25.1\\
		& $\varepsilon$ & 0.45 & 0.46 & 0.48 & 0.46 & \textbf{0.44} & 0.45 &\\
		& & & & & & & &\\
		$U_{Grav}$ & $\alpha$ & 1.20 & 1.53 & 3.53 & 0.82 & 1.02 & \textbf{0.69} & 11.2\\
		& $\varepsilon$ & 0.42 & 0.45 & 0.42 & 0.40 & 0.40 & \textbf{0.40} &\\
		& & & & & & & &\\ \htw
	\end{tabular}
	\caption{Comparison of the results obtained from the regressions and those obtained with the model described above only including early-types with the covariance matrix of \citet{Bernardi03b}. Each row represents the model prediction assuming as fundamental variable that in the leftmost columnn. $\alpha$'s and $\varepsilon$'s are slopes and intrinsic scatters, respectively. The rightmost column shows the $\chi^2$ values. The values in bold are those obtained through linear regressions.} 
	\label{modellino_mine}
\end{table}  

\begin{table}  
	\centering
	\setlength{\tabcolsep}{3pt}
	\begin{tabular}{|c|c c c c c c c c|}
		\hdr
	 Fund. / Obs. & & $L$ & $R_e$ & $V$ & $M_{Hop}$ & $M_{Vir}$ & $U_{Grav}$ & $\chi^2$ \\ \hline    
		$L$ & $\alpha$ & \textbf{1.00} & 1.18 & 2.27 & 0.57 & 0.77 & 0.50 & 86.4 \\
		& $\varepsilon$ & \textbf{0.47} & 0.50 & 0.51 & 0.50 & 0.48 & 0.48 &\\
		& & & & & & & &\\
		$R_e$ & $\alpha$ & 0.57 & \textbf{0.91} & 1.05 & 0.33 & 0.50 & 0.31 & 293.7\\
		& $\varepsilon$ & 0.62 & \textbf{0.60} & 0.64 & 0.63 & 0.62 & 0.62 &\\
		& & & & & & & &\\
		$V$ & $\alpha$ & 1.12 & 1.08 & \textbf{4.48} & 0.93 & 0.96 & 0.71 & 14.6\\
		& $\varepsilon$ & 0.48 & 0.55 & \textbf{0.38} & 0.40 & 0.47 & 0.43 &\\
		& & & & & & & &\\
		$M_{Hop}$ & $\alpha$ & 1.18 & 1.40 & 3.89 & \textbf{0.88} & 1.01 & 0.71 & 37.2\\
		& $\varepsilon$ & 0.40 & 0.45 & 0.36 & \textbf{0.35} & 0.38 & 0.36 &\\
		& & & & & & & &\\
		$M_{Vir}$ & $\alpha$ & 0.95 & 1.28 & 2.39 & 0.60 & \textbf{0.80} & 0.53 & 76.7\\
		& $\varepsilon$ & 0.46 & 0.47 & 0.49 & 0.47 & \textbf{0.45} & 0.46 &\\
		& & & & & & & &\\
		$U_{Grav}$ & $\alpha$ & 1.10 & 1.41 & 3.15 & 0.75 & 0.94 & \textbf{0.63} & 42.5\\
		& $\varepsilon$ & 0.40 & 0.43 & 0.42 & 0.39 & 0.39 & \textbf{0.39} &\\
		& & & & & & & &\\ \htw
	\end{tabular}
	\caption{Same as Tab.\tild\ref{modellino_mine} also including bulges coming from decompositions of spirals.} 
	\label{modellino_mine_clbul}
\end{table}

\subsection{The importance of the covariance matrix} \label{par44}
We now show the critical dependence on the covariance matrix of the results we obtained in the last section. We have in fact used a covariance matrix from the SDSS sample of early type galaxies that is more homogeneous and larger than ours which is also biased towards more luminous objects (\cite{Shankar16} and Fig.\tild\ref{debias}). Thus, we show the consequences of using, at least, the variances obtained from our sample through a maximum likelihood analysis such as the one presented in the last section\footnote{Note that, given the small number of data-points, the estimate of the whole covariance matrix from our sample could lead to huge errors.}.\\
\indent We start by considering all classical bulges of our sample. Taking the correlation coefficients from Tab.\tild\ref{coeffs} and fitting the three variances using eq.\tild\ref{trivariate}, we get the covariance matrix reported in Tab.\tild\ref{coeffs_clbul}, which leads to the results for our analysis of Tab.\tild\ref{modellino_mine_clbul_ourvar}. \\
This shows how critical the choice of the covariance matrix turns out to be. In fact, using the variances from our sample, the BHFP has a $\chi^2$ slightly lower than that of $V$. When instead we just use the early-types of our sample, the covariance matrix (Tab.\tild\ref{coeffs_ETs}) leads to the results reported in Tab.\tild\ref{modellino_mine_ourvar}. \\
\indent In this last case, both the BHFP and $U_{grav}$ reproduce \textit{almost perfectly} the other relations, while $V$ has a higher $\chi^2$ than all its linear combination with $R_e$. This can be explained by the fact that the Kormendy relation we derived for our sample has a much lower intrinsic scatter than in other works on this topic (e.g. \citealt{S16}). The quantity $V$ alone yields the worst predictions for the monovariate correlations \mbhnosp-$L$ and \mbhnosp-\renosp, while this estimate improves drastically when $V$ gets combined with $R_e$. Notice that despite the reduced number of data-points (49 for the early-type subsample) the errors in the variance estimates reassure us about the robustness of the minimization results. Thus, we caution that using a covariance matrix estimated from a biased and heterogeneous sample can significantly alter the results of the analysis.

\begin{table}
	\centering
	\begin{tabular}{|c c|}
		\hdr 
		Coefficient & Value \\ \hline
	    $\sigma_V$ & 0.17 $\pm$ 0.01 \\
		$\sigma_L$ & 0.64 $\pm$ 0.03 \\
		$\sigma_{R_e}$ & 0.55 $\pm$ 0.03 \\ \htw 
	\end{tabular}
	\caption{Variances obtained from our classic bulge sample by fitting eq.\tild\ref{trivariate} to it and taking correlation coefficients from Tab.\tild\ref{coeffs}.}
	\label{coeffs_clbul}
\end{table}

\begin{table}  
	\centering
	\setlength{\tabcolsep}{3pt}
	\begin{tabular}{|c|c c c c c c c c|}
		\hdr
	 Fund. / Obs. & & $L$ & $R_e$ & $V$ & $M_{Hop}$ & $M_{Vir}$ & $U_{Grav}$ & $\chi^2$ \\ \hline    
		$L$ & $\alpha$ & \textbf{1.00} & 1.03 & 2.93 & 0.69 & 0.79 & 0.56 & 39.9 \\
		& $\varepsilon$ & \textbf{0.47} & 0.55 & 0.61 & 0.55 & 0.49 & 0.51 &\\
		& & & & & & & &\\
		$R_e$ & $\alpha$ & 0.67 & \textbf{0.91} & 1.55 & 0.45 & 0.58 & 0.39 & 203.3\\
		& $\varepsilon$ & 0.65 & \textbf{0.60} & 0.73 & 0.68 & 0.62 & 0.65 &\\
		& & & & & & & &\\
		$V$ & $\alpha$ & 0.89 & 0.73 & \textbf{4.48} & 0.87 & 0.76 & 0.61 & 3.3\\
		& $\varepsilon$ & 0.59 & 0.72 & \textbf{0.38} & 0.42 & 0.57 & 0.49 &\\
		& & & & & & & &\\
		$M_{Hop}$ & $\alpha$ & 1.00 & 1.02 & 4.18 & \textbf{0.89} & 0.86 & 0.66 & 2.5\\
		& $\varepsilon$ & 0.48 & 0.59 & 0.39 & \textbf{0.35} & 0.44 & 0.38 &\\
		& & & & & & & &\\
		$M_{Vir}$ & $\alpha$ & 0.95 & 1.09 & 3.03 & 0.71 & \textbf{0.80} & 0.58 & 33.9\\
		& $\varepsilon$ & 0.48 & 0.50 & 0.58 & 0.51 & \textbf{0.45} & 0.47 &\\
		& & & & & & & &\\
		$U_{Grav}$ & $\alpha$ & 1.02 & 1.11 & 3.66 & 0.82 & 0.86 & \textbf{0.63} & 10.7\\
		& $\varepsilon$ & 0.44 & 0.51 & 0.49 & 0.41 & 0.40 & \textbf{0.39} &\\
		& & & & & & & &\\ \htw
	\end{tabular}
	\caption{Same as Tab.\tild\ref{modellino_mine_clbul} using variances directly derived from our sample.} 
	\label{modellino_mine_clbul_ourvar}
\end{table}   

\begin{table}
	\centering
	\begin{tabular}{|c c|}
		\hdr 
		Coefficient & Value \\ \hline
		$\sigma_V$ & 0.16 $\pm$ 0.01 \\
		$\sigma_L$ & 0.58 $\pm$ 0.04 \\
		$\sigma_{R_e}$ & 0.46 $\pm$ 0.03 \\ \htw 
	\end{tabular}
	\caption{Same as Tab.\tild\ref{coeffs_clbul} only including early-types.}
	\label{coeffs_ETs}
\end{table}

\begin{table}  
	\centering
	\setlength{\tabcolsep}{3pt}
	\begin{tabular}{|c|c c c c c c c c|}
		\hdr
	 Fund. / Obs. & & $L$ & $R_e$ & $V$ & $M_{Hop}$ & $M_{Vir}$ & $U_{Grav}$ & $\chi^2$ \\ \hline    
		$L$ & $\alpha$ & \textbf{1.13} & 1.25 & 3.21 & 0.79 & 0.92 & 0.64 & 11.6 \\
		& $\varepsilon$ & \textbf{0.44} & 0.54 & 0.60 & 0.52 & 0.47 & 0.48 &\\
		& & & & & & & &\\
		$R_e$ & $\alpha$ & 0.81 & \textbf{1.14} & 1.80 & 0.54 & 0.72 & 0.48 & 88.5\\
		& $\varepsilon$ & 0.64 & \textbf{0.59} & 0.74 & 0.68 & 0.62 & 0.65 &\\
		& & & & & & & &\\
		$V$ & $\alpha$ & 0.93 & 0.81 & \textbf{4.32} & 0.90 & 0.86 & 0.66 & 7.7\\
		& $\varepsilon$ & 0.59 & 0.71 & \textbf{0.41} & 0.43 & 0.54 & 0.48 &\\
		& & & & & & & &\\
		$M_{Hop}$ & $\alpha$ & 1.09 & 1.17 & 4.30 & \textbf{0.92} & 0.95 & 0.71 & 0.2\\
		& $\varepsilon$ & 0.48 & 0.58 & 0.40 & \textbf{0.38} & 0.45 & 0.40 &\\
		& & & & & & & &\\
		$M_{Vir}$ & $\alpha$ & 1.07 & 1.30 & 3.48 & 0.80 & \textbf{0.93} & 0.65 & 8.4\\
		& $\varepsilon$ & 0.47 & 0.49 & 0.55 & 0.49 & \textbf{0.44} & 0.45 &\\
		& & & & & & & &\\
		$U_{Grav}$ & $\alpha$ & 1.11 & 1.27 & 3.94 & 0.88 & 0.96 & \textbf{0.69} & 2.0\\
		& $\varepsilon$ & 0.45 & 0.51 & 0.47 & 0.42 & 0.41 & \textbf{0.40} &\\
		& & & & & & & &\\ \htw
	\end{tabular}
	\caption{Same as Tab.\tild\ref{modellino_mine_clbul_ourvar} using early-types only.} 
	\label{modellino_mine_ourvar} 
\end{table}

\section{Conclusions}
We have studied the scaling relations between SMBHs and their host galaxies, extending our analysis to the four-dimensional case. In this work, we analytically combine for the first time the whole FP with BH masses deriving formulae to express slopes and intrinsic scatter of the BH-hosts as functions of the covariance matrix. Conversely to the findings of \citet{Hopkins07b}, the fundamental scaling relation seems to be the canonical \mbhnosp-\sinosp, even though a bivariate relation $\mbhmath \propto \simath^{\sim 4} \remath^{\sim\beta}$ with an exponent $0.4 \leq \beta \leq 1$ acceptably explains the other correlations when combined with the FP. We have seen that this result is independent of whether we include or not classical bulge parameters coming from decompositions (but the same also holds for pseudobulges, see App.\tild\ref{Appendice2}) but also that it critically depends on the covariance matrix one chooses out for the analysis.\\ 
\indent In the only other work where the whole FP is taken into account \citep{VDB16} the main conclusion also points to a BH-host coevolution driven by \si solely. Indeed, the intrinsic scatter of the \mbhnosp-\si relation is not significantly improved by higher dimensionality (Sec.\tild\ref{regr}). In the work of \citet{VDB16}, no improvement at all is found, but in that work the focus is on whole galaxies rather than decomposition, and BHs are known to correlate poorly (if at all) with disk parameters \citet{Kormendy11}.\\
\indent Provided that each galaxies hosts, or has hosted, an AGN at its center \citep{Soltan82}, we believe that the \mbhnosp-\si relation is established when \mbh reaches a critical value $M_\sigma$ proportional to $\simath^4$ which signals a change of the AGN feedback from momentum-driven to energy-driven \citep{King03, King05}. While in the first case the efficient Compton cooling enables SMBHs and bulges to coevolve pacifically, in the second case the energy output from BH winds is two orders of magnitude larger than the bulge's binding energy, thus seriously threatening the integrity of the host spheroid \citep{King15}. However, the simple $M_\sigma \propto \simath^4$ dependence arises from the (unrealistic) description of a galaxy as a singular isothermal sphere (SOI, see eqs. 37-41 of \citealt{King15}). If the potential has a more complicated form, then a dependence on $R_e$ might come out, but the whole picture is still uncertain. Furthermore, since the modeling of the FP as a trivariate gaussian \citep{Bernardi03b} introduces covariances and correlations between the bulge parameters and considering how tight this relation is, then it can be expected that a bivariate correlation can provide acceptable results. Besides, as we have seen in Sec.\tild\ref{par44}, the results differ depending on the variances, so the whole picture is still uncertain.\\
\indent These problems could be resolved by the future development of new generation facilities. For instance, the sample used in this work is very heterogeneous regardless of the variable we consider. This is of particular relevance for BH mass estimates, which challenge current-day facilities and often give disagreeing results when different techniques are applied to the same galaxy \citep{K13}, and, furthermore, the subset of BH masses nowadays available are likely to be a biased-high subsample \citep{Shankar16, Shankar17, Shankar19}. Velocity dispersions should also be measured with the same instrument, but such coverage is not available. As far as concerns photometry, decomposing all spirals of our sample ($\sim 60$) using $K$-band data ($Spitzer$ data are not available for the whole sample) steps beyond the purposes of this work, and more accurate multi-component decompositions could lead to significantly different results (see e.g. the latest results of \citealt{Davis18, Davis19, Sahu19}). \\
We finally note that the fact that we \textit{assume} each relation to be the fundamental one in order to see how it performs in predicting the others is itself an approximation. We should use the whole eqs.\tild\ref{L}, \ref{R} and \ref{V} to quantify the correction introduced by the FP on each of the monovariate scaling relations.

\section*{Acknowledgements}
We thank the anonymous referee for comments on the manuscript. GL acknowledges support from the European Union’s Horizon 2020 Sundial Innovative Training Network, grant n.721463.
This work has made use of the HyperLeda database \citep{Paturel03}.

\bibliographystyle{mnras}
\bibliography{bibl}

\appendix
\section{Notes on discarded galaxies} \label{Appendice}
\textit{Galaxies omitted because of incomplete data.} The following objects were omitted because of the unavailability of photometric measurements in the infrared: \textit{Milky Way}\footnote{For our own Galaxy, a $K$-band magnitude value can be found in \citet{K13} whom, however, do not provide effective radii.}, \textit{NGC1194}, \textit{NGC4526}, \textit{NGC6264}, \textit{NGC6323}.\\
\textit{NGC3607:} We follow \citet{K13} in omitting this galaxy because, in this case, not allowing for a DM halo yields underestimated results because the BH SOI is not well resolved.\\
\textit{NGC4388:} The \mbh value is uncertain because of the lack of a systemic maser in this galaxy and because the rotation might even be non-Keplerian \citet{Kuo11}. The velocity dispersion is uncertain since it neglects rotation inside the effective radius of the pseudobulge (see \cite{K13}). Moreover, \citet{Savorgnan16a} report complications due to dust absorption when deriving the photometry.\\
\textit{NGC2974, NGC3414, NGC4552 (M89), NGC4621 (M59), NGC5813, NGC5846:} These (uncertain) \mbhnosp's come from an unlabeled plot from \citet{Cappellari08}, who do not provide any information or details about the observations and data analysis. We thus consider the BH masses unreliable.\\
\textit{NGC3079:} We reject the BH mass value because the rotation curve is flat \citealt{Kondratko05}. \citet{S16} accept the galaxy stating that the estimate agrees with \citet{Yamauchi04}. However, both authors do not provide an exact \mbh value. This galaxy also shows a steep drop in \si in the central regions, probably because of bar streaming motions \citep{Graham13}.\\
\textit{NGC4486B:} We follow \citet{S16} who omit the galaxy because of the abnormally large BH mass. Indeed, this galaxy is well known to be an outlier \citep{Gultekin09b}. Moreover, the \mbh value was derived without allowing for a DM halo and comes from unpublished literature \citep{S16}.\\
\textit{NGC4736, NGC4826:} We reject these (commonly accepted) \mbh estimates because details on the observation are not provided (see \citealt{Kormendy11}).\\
\textit{NGC1300:} This galaxy appears to have an uncertain velocity dispersion. In fact, the value from \citet{S16} is much lower than that reported in the sample of \citet{VDB16}. Using the first value would make this galaxy the largest outlier in the \mbhnosp-\si relation, while, interestingly, \citet{VDB16} reports the second value to be too high.\\
\textit{NGC2787:} This galaxy hosts both a classical and a pseudobulge. Since it is unclear what \citet{Savorgnan16a} model in their analysis, even if the pseudobulge seems to be more prominent \citep{K13}, we do not trust the photometry and omit this galaxy.\\
\textit{NGC2960:} We omit this galaxy because of its extremely uncertain morphology. In fact, \citet{S16} classify the galaxy as an E2 but give it a $T$ flag value of 3, which would make this galaxy a \textit{pure pseudobulge}.\\
\textit{IC1481:} We follow \citet{S16} in omitting this galaxy because of a merger in progress which prevents the determination of a reliable photometric profile. 

\section{Results including pseudobulges} \label{Appendice2} 
This appendix contains the results of the analysis described in Sec.\tild\ref{modellino} including pseudobulges, which do not seem to correlate with their SMBHs (cfr. §\tild\ref{Sec.regression_results}). The covariance matrix for this case is reported in Tab.\tild\ref{coeffs_all}. By computing the analogous of Tab.\tild\ref{modellino_mine} (Tab.\tild\ref{last}) and of Tab.\tild\ref{modellino_mine_ourvar} (Tab.\tild\ref{modellino_mine_all_ourvar}) we see that, although in this case the usage of the covariance matrix is not so well justified as pseudobulges form stars and thus have a much broader range of colors than classical bulges, no significant differences are found.

\begin{table}
	\centering
	\begin{tabular}{|c c|}
		\hdr 
		Coefficient & Value \\ \hline
		$\sigma_V$ & 0.18 $\pm$ 0.01 \\
		$\sigma_L$ & 0.69 $\pm$ 0.03 \\
		$\sigma_{R_e}$ & 0.64 $\pm$ 0.04 \\ \htw  
	\end{tabular}
	\caption{Same as Tab.\tild\ref{coeffs_clbul} using the full sample.}
	\label{coeffs_all}
\end{table}

\begin{table} 
	\centering
	\setlength{\tabcolsep}{4pt}
	\begin{tabular}{|c|c c c c c c c c|}
		\hdr
	 Fund. / Obs. & & $L$ & $R_e$ & $V$ & $M_{Hop}$ & $M_{Vir}$ & $U_{Grav}$ & $\chi^2$ \\ \hline    
		$L$ & $\alpha$ & \textbf{1.12} & 1.39 & 2.67 & 0.68 & 0.89 & 0.59 & 118.1 \\
		& $\varepsilon$ & \textbf{0.51} & 0.54 & 0.56 & 0.54 & 0.52 & 0.53 &\\
		& & & & & & & &\\
		$R_e$ & $\alpha$ & 0.69 & \textbf{1.07} & 1.28 & 0.43 & 0.62 & 0.39 & 413.4\\
		& $\varepsilon$ & 0.63 & \textbf{0.63} & 0.66 & 0.64 & 0.62 & 0.63 &\\
		& & & & & & & &\\
		$V$ & $\alpha$ & 1.26 & 1.22 & \textbf{5.07} & 1.05 & 1.14 & 0.82 & 41.8\\
		& $\varepsilon$ & 0.53 & 0.63 & \textbf{0.42} & 0.43 & 0.49 & 0.46 &\\
		& & & & & & & &\\
		$M_{Hop}$ & $\alpha$ & 1.31 & 1.68 & 4.28 & \textbf{0.99} & 1.13 & 0.79 & 70.6\\
		& $\varepsilon$ & 0.42 & 0.45 & 0.38 & \textbf{0.37} & 0.39 & 0.37 &\\
		& & & & & & & &\\
		$M_{Vir}$ & $\alpha$ & 1.03 & 1.43 & 2.78 & 0.68 & \textbf{0.88} & 0.58 & 117.4\\
		& $\varepsilon$ & 0.47 & 0.46 & 0.49 & 0.47 & \textbf{0.46} & 0.46 &\\
		& & & & & & & &\\
		$U_{Grav}$ & $\alpha$ & 1.18 & 1.59 & 3.48 & 0.82 & 1.02 & \textbf{0.69} & 66.2\\
		& $\varepsilon$ & 0.42 & 0.43 & 0.43 & 0.40 & 0.40 & \textbf{0.41} &\\
		& & & & & & & &\\ \htw
	\end{tabular}
	\caption{Same as Tab.\tild\ref{modellino_mine_all_ourvar} using the covariance matrix reported in Tab.\tild\ref{coeffs}.} 
	\label{last}
\end{table}

\begin{table} 
	\centering
	\setlength{\tabcolsep}{4pt}
	\begin{tabular}{|c|c c c c c c c c|}
		\hdr
	 Fund. / Obs. & & $L$ & $R_e$ & $V$ & $M_{Hop}$ & $M_{Vir}$ & $U_{Grav}$ & $\chi^2$ \\ \hline    
		$L$ & $\alpha$ & \textbf{1.12} & 1.08 & 3.33 & 0.78 & 0.84 & 0.61 & 49.7 \\
		& $\varepsilon$ & \textbf{0.53} & 0.63 & 0.71 & 0.61 & 0.56 & 0.57 &\\
		& & & & & & & &\\
		$R_e$ & $\alpha$ & 0.90 & \textbf{1.07} & 2.08 & 0.65 & 0.76 & 0.53 & 189.4\\
		& $\varepsilon$ & 0.70 & \textbf{0.63} & 0.86 & 0.73 & 0.65 & 0.68 &\\
		& & & & & & & &\\
		$V$ & $\alpha$ & 1.01 & 0.76 & \textbf{5.07} & 0.97 & 0.86 & 0.68 & 11.4\\
		& $\varepsilon$ & 0.69 & 0.86 & \textbf{0.42} & 0.45 & 0.61 & 0.53 &\\
		& & & & & & & &\\
		$M_{Hop}$ & $\alpha$ & 1.14 & 1.13 & 4.64 & \textbf{0.99} & 0.93 & 0.71 & 1.7\\
		& $\varepsilon$ & 0.52 & 0.60 & 0.40 & \textbf{0.37} & 0.45 & 0.39 &\\
		& & & & & & & &\\
		$M_{Vir}$ & $\alpha$ & 1.10 & 1.18 & 3.68 & 0.83 & \textbf{0.88} & 0.64 & 27.8\\
		& $\varepsilon$ & 0.51 & 0.50 & 0.61 & 0.52 & \textbf{0.46} & 0.47 &\\
		& & & & & & & &\\
		$U_{Grav}$ & $\alpha$ & 1.14 & 1.19 & 4.18 & 0.91 & 0.92 & \textbf{0.69} & 8.9\\
		& $\varepsilon$ & 0.47 & 0.51 & 0.50 & 0.42 & 0.41 & \textbf{0.41} &\\
		& & & & & & & &\\ \htw
	\end{tabular}
	\caption{Same as Tab.\tild\ref{modellino_mine_clbul_ourvar} using the full sample and the variances of Tab.\tild\ref{coeffs_all}.} 
	\label{modellino_mine_all_ourvar}
\end{table}

\bsp	
\label{lastpage}
\end{document}